\documentclass[
 table,
 twocolumn,
 superscriptaddress,
 amsmath,amssymb,
 aps,
 pra,
]{revtex4-2}

\usepackage[utf8]{inputenc}
\usepackage[T1]{fontenc}
\usepackage{amssymb}
\usepackage{mathtools}
\usepackage{bbold}
\usepackage{bm}
\usepackage[position=top, caption=false]{subfig}
\usepackage[table,x11names,dvipsnames,table]{xcolor}
\usepackage{tikz}
\usetikzlibrary {shapes.geometric}

\usepackage{amsthm}
\usepackage{amsmath}
\theoremstyle{definition}
\newtheorem{definition}{Definition}

\newtheorem{exmp}{Example}[section]

\DeclareMathOperator*{\argmin}{arg\,min}

\definecolor{car_red}{HTML}{FF244E}
\definecolor{car_blue}{HTML}{2196F3}
\definecolor{dark_grey}{HTML}{303030}
\begin{document}

\newcommand\circlesymbol{\,\begin{tikzpicture}
    \node[fill, shape=circle, inner sep=0pt, minimum size=8pt, draw] {};
    \end{tikzpicture}\,}
\newcommand\trianglesymbol{\,\begin{tikzpicture}
    \node[fill, regular polygon, regular polygon sides=3, inner sep=0pt, minimum size=10pt, draw] {};
    \end{tikzpicture}\,}
\newcommand\squaresymbol{\,\begin{tikzpicture}
    \node[fill, regular polygon, regular polygon sides=4, inner sep=0pt, minimum size=10pt, draw] {};
    \end{tikzpicture}\,}
\newcommand\pentagonsymbol{\,\begin{tikzpicture}
    \node[fill, regular polygon, regular polygon sides=5, inner sep=0pt, minimum size=9pt, draw] {};
    \end{tikzpicture}\,}

\title{Optimisation-Free Recursive QAOA for the Binary Paint Shop Problem}

\affiliation{School of Physics, University of Melbourne, VIC, Parkville, 3010, Australia.}
\affiliation{School of Mathematics and Statistics, University of Melbourne, VIC, Parkville, 3010, Australia.}

\author{Gary J Mooney}
\email{mooneyg@unimelb.edu.au}
\affiliation{School of Physics, University of Melbourne, VIC, Parkville, 3010, Australia.}
\author{Jedwin Villanueva}
\affiliation{School of Physics, University of Melbourne, VIC, Parkville, 3010, Australia.}
\author{Bhaskar Roy Bardhan}
\affiliation{Research and Advanced Engineering, Ford Motor Company, Dearborn, MI 48124, USA.}
\author{Joydip Ghosh}
\affiliation{Research and Advanced Engineering, Ford Motor Company, Dearborn, MI 48124, USA.}
\author{Charles D Hill}
\affiliation{School of Physics, University of Melbourne, VIC, Parkville, 3010, Australia.}
\affiliation{School of Mathematics and Statistics, University of Melbourne, VIC, Parkville, 3010, Australia.}
\author{Lloyd C L Hollenberg}
\email{lloydch@unimelb.edu.au}
\affiliation{School of Physics, University of Melbourne, VIC, Parkville, 3010, Australia.}

\date{\today}

\begin{abstract}
The Quantum Approximate Optimisation Algorithm (QAOA) is a leading candidate for near-term quantum advantage, yet its practical impact is hindered by limited performance on symmetric local Hamiltonians and the costly optimisation of variational parameters. The Recursive-QAOA (RQAOA) introduced by Bravyi et al. Phys. Rev.
Lett. \textbf{125}, 260505 (2020), addresses the first limitation while also reducing circuit size, and parameter transfer techniques can be used to effectively bypass the optimisation loop. In this work, we combine these two ideas to develop an optimisation-free RQAOA and evaluate its performance on the Binary Paint Shop Problem (BPSP)---an optimisation problem found in manufacturing where a sequence of cars must be painted under constraints while minimising the number of colour changes. The BPSP can be formulated as an Ising ground state problem with a symmetric local Hamiltonian in the form of MAX-CUT and properties well-suited for the application of QAOA parameter transfer. We benchmark QAOA and RQAOA with parameter transfer against classical solvers and heuristics, and investigate their resilience to suboptimal parameters. For circuit optimisation, we use reverse causal cones (RCC) and introduce a method of trimming outer two-qubit gates. To estimate the classical resources needed to simulate these quantum algorithms, we compute entanglement entropy and bond dimensions using matrix product state methods. We also compare circuit sizes and measurement counts across implementations. Our results show that RQAOA is inherently robust to parameter deviations, maintaining near-optimal solutions without noticeable degradation under parameter transfer while substantially reducing quantum resource requirements compared to QAOA. This highlights a viable route toward scalable quantum optimisation without the overhead of the classical optimisation loop and its challenges with barren plateaus.
\end{abstract}

\maketitle

\section{Introduction} \label{sec:introduction}
With quantum computing technology rapidly advancing within the noisy intermediate scale quantum (NISQ) era~\cite{preskill2018quantum}, an important research question receiving a lot of attention is what real-world problems can quantum computing be advantageous for in the near future. A promising candidate for this endeavour is optimisation. Optimisation problems are generally broadly applicable and often difficult to solve exactly, in many cases they can be efficiently mapped to quantum algorithms~\cite{Brandhofer2022, Khumalo2022, PhysRevA.104.012403, Barkoutsos_2020}. A common quantum algorithmic approach to optimisation is the quantum approximate optimisation algorithm (QAOA)~\cite{farhi2014quantum, blekos2024review, abbas2024challenges}. It is a popular quantum-classical hybrid algorithm that can find good approximations to the ground states of Ising Hamiltonians that are encoding the solutions to a quadratic unconstrained binary optimisation (QUBO) problem instance. The QAOA circuits are relatively short compared to mainstay circuit model algorithms like Shor's~\cite{shor1994algorithms} and Grover's~\cite{grover1996fast} algorithms, making them well-suited for analysis on NISQ computing. It remains to be seen whether general quantum optimisation approaches, such as QAOA, can compete with classical heuristics that have been tailored over many decades to specific classes of optimisation problems like the travelling salesman problem (TSP)~\cite{johnson2007experimental, Applegate2003ChainedLF, HELSGAUN2000106, 10.1145/290179.290180, johnson_mcgeoch_1995, 10.1007/3-540-52846-6_97}. The recursive quantum approximate optimisation algorithm (RQAOA) introduced by Bravyi et al.~\cite{bravyi2020obstacles} has shown potential. Motivated by alleviating inherent computational limits present in QAOA with symmetric local Hamiltonians, it produces outstanding improvements in solution qualities over baseline QAOA. In addition, it provides a trade-off between quantum circuit resources and number of circuits by reducing the circuit to reverse-causal cones (RCCs) which calculate $\langle ZZ \rangle$ correlations without needing to generate the full QAOA ansatz state. However, a noticeable drawback in RQAOA is that it uses QAOA as a subroutine, executing it for the initial problem graph and subsequent graphs generated by incrementally reducing its size until it is small enough to classically solve. Each application of QAOA involves optimising its parameters, which often requires hundreds of circuits for even modest problem sizes and QAOA depths while also suffering from barren plateaus~\cite{holmes2022connecting, wang2021noise, mcclean2018barren}.

In this work, we demonstrate that combining RQAOA with precomputed QAOA parameters overcomes these challenges, achieving near-optimal performance while remaining robust to parameter noise. This robustness suggests that parameter tuning for RQAOA is scalable, providing strong solutions without extensive parameter optimisation.

The binary paint shop problem (BPSP) is a combinatorial optimisation challenge important both economically and environmentally to the automotive industry, with the goal to paint sequences of cars under certain constraints while minimising the number of times the colour needs to be swapped~\cite{zhang2018environment}. It has a simpler formulation than the general paint shop problem (PSP) used in actual car factories, however it is still NP-complete and APX-hard, making it an appealing problem for foundational studies~\cite{bonsma2006complexity, meunier2009paintshop}. The BPSP instances can conveniently be mapped efficiently to symmetric Ising graphs that approach a regular structure as problem sizes increase. Specifically, the problem Ising graphs approach 4-regular with unit-couplings ($\pm 1$ weights), which is equivalent to MAX-CUT~\cite{garey1974some} on a 4-regular graph with $\pm1$ weighted edges, a ubiquitous and well-studied problem in theoretical computer science. Under a fixed-angle conjecture, this regularity is well suited for the use of precomputation techniques to find good estimates to the optimised QAOA parameters, typically well within~1\% approximation ratio on equivalent MAX-CUT problems~\cite{wurtz2021fixed, ozaeta2022expectation}. Previous work on BPSPs used precomputed parameters to calculate QAOA depth 1 and 2 expectations on problems of up 100 qubits~\cite{streif2021beating, streif2020training}. For graphs with irregular structure, the median optimal parameters calculated from many smaller instances can be transferred to unseen larger instances using the QAOAKit Python library~\cite{shaydulin2021qaoakit, shaydulin2023parameter}.

Building on this motivation, we introduce and study RQAOA in combination with precomputed parameters as an approach to solve the BPSP. We benchmark the solution qualities obtained by precomputing and optimising QAOA and RQAOA against classical heuristics and exact solvers, comparing both fixed angle and QAOAKit methods of obtaining precomputed parameters for QAOA and for each recursive step of RQAOA. We demonstrate the substantial improvement that RQAOA provides over QAOA, producing near-optimal solutions on generated BPSP instances up to 20 car bodies (mapped to 20 qubits). We find that RQAOA with precomputed parameters obtained from QAOAKit perform effectively as good as optimal. Precomputed fixed-angle parameters appears to still perform well even though the problem graph regularity is compromised during recursive reductions in RQAOA. Additionally, we observe that RQAOA exhibits a level of noise robustness in QAOA parameters, and that the typical displacements between QAOAKit and optimal parameters fall within this regime. This highlights RQAOA combined with precomputed parameters as a powerful yet scalable approach for BPSP and likely non-BPSP Ising graphs as well. Motivated by the question of algorithm scalability, we also investigate the classical and quantum resource requirements for RQAOA. In particular we obtain quantum circuit metrics like CNOT counts, CNOT depths and qubit counts, as well as circuit counts and matrix product state (MPS) entanglement entropies and bond dimensions for varying problem sizes and QAOA depths. We consider the circuits and their RCCs, and after compilation onto a 27-qubit IBM Quantum device layout.

Shortly after this paper was uploaded to the arXiv, another study by Vijendran et al.~\cite{vijendran2025classical} also applied RQAOA to the BPSP. They observed that as the problem size increases, the performance of RQAOA at $p=1$ with optimised parameters gradually declines. Taken together with our finding that RQAOA appears robust to parameter noise, this suggests that while RQAOA with $p=1$ performs strongly on BPSP instances at modest scales (around 1000 car bodies), understanding how to maintain this performance for increasingly larger systems remains an open and interesting question.

\section{Methods}

\subsection{The Quantum Approximate Optimisation Algorithm (QAOA)}

\begin{figure}
     \centering
     \includegraphics[width=\linewidth]{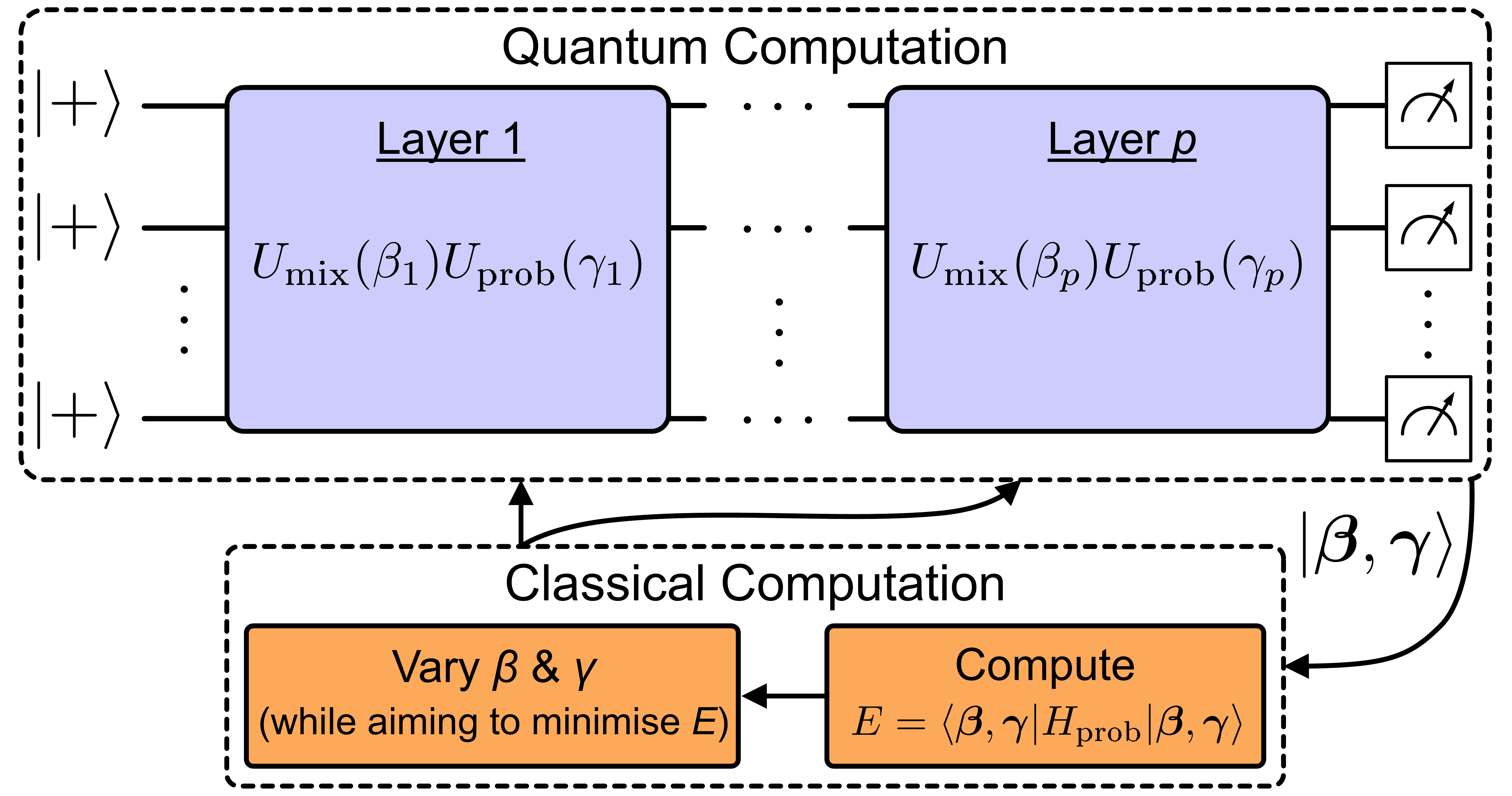}
     \caption[]{A diagram of the quantum approximate optimisation algorithm (QAOA) and its quantum-classical hybrid nature. The ansatz state $|\boldsymbol{\beta}, \boldsymbol{\gamma} \rangle$ is prepared by running the ansatz circuit with an initial selection of parameters $\boldsymbol{\beta}$ and $\boldsymbol{\gamma}$. The measurement results of the ansatz circuit are used to classically calculate the Hamiltonian $H_\text{prob}$ energy corresponding to the problem encoding, which acts as the objective function for optimisation. The energy calculation guides the adjustment of parameters $\boldsymbol{\beta}$ and $\boldsymbol{\gamma}$ based on the optimisation strategy, leading to the preparation of a new ansatz circuit. This process is repeated until the parameters $\boldsymbol{\beta}$ and $\boldsymbol{\gamma}$ have been optimised, at which point the corresponding ansatz state $|\boldsymbol{\beta}, \boldsymbol{\gamma} \rangle$ can be measured to provide approximate solutions to the target QUBO problem.
     \label{fig:qaoa_diagrams}}
\end{figure}
Here we give a brief overview of QAOA. For deeper insight into the algorithm and how it's implemented, we include a lengthy derivation and further details in App.~\ref{sec:qaoa-derivation}.

The QAOA~\cite{farhi2014quantum}, shown in Fig.~\ref{fig:qaoa_diagrams}, is a form of variational quantum eigensolver (VQE)~\cite{peruzzo2014variational} based on the variational principle of quantum mechanics. It is a quantum-classical hybrid algorithm for approximating the ground state of a problem Ising Hamiltonian $H_\text{prob}$ (where we will be using qubits in place of spins) that encodes solutions to a quadratic unconstrained binary optimisation (QUBO) problem instance. A parameterised QAOA ansatz state is constructed from information about the problem Hamiltonian 
\begin{align}
    |\boldsymbol{\beta}, \boldsymbol{\gamma}\rangle &:= U(\boldsymbol{\beta}, \boldsymbol{\gamma})|+\rangle^{\otimes N}\\
    &= \prod_{l=1}^p U_\text{mix}(\beta_l) U_\text{prob}(\gamma_l)|+\rangle^{\otimes N},
\end{align}
where $p$ is the number of QAOA layers, $\boldsymbol{\beta}:=\{\beta_1, \beta_2, \ldots \beta_p\}$ and $\boldsymbol{\gamma}:=\{\gamma_1, \gamma_2, \ldots \gamma_p\}$ are the QAOA parameters, $U_\text{mix}(\beta_i)$ is the mixer operator, and $U_\text{prob}(\gamma_i)$ is the phase operator encoding the problem Hamiltonian. By tuning the QAOA parameters to minimise the energy 
\begin{equation}
    E = \langle \boldsymbol{\beta}, \boldsymbol{\gamma} | H_\text{prob} |\boldsymbol{\beta}, \boldsymbol{\gamma}\rangle,
\end{equation}
we obtain a state with an energy close to that of the ground state. The lowest energy is desired because the ansatz state energy is always an upper bound to the ground state energy due to the variational principle of quantum mechanics. 

\subsection{The Binary Paint Shop Problem (BPSP)}

\begin{figure*}
     \centering
     \includegraphics[width=0.75\linewidth]{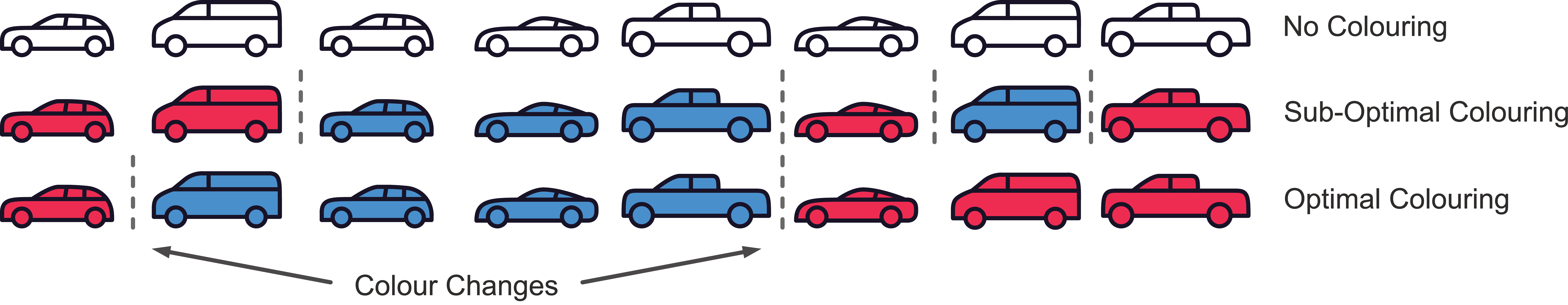}
     \caption[]{An example of the binary paint shop problem (BPSP) consisting of a sequence of 8 car instances and 4 body types. There are two cars for each body type and they must be painted different colours from within the selection of two available colours (red and blue in this case). The aim is to assign colours in a way that minimises the number of colour changes in the sequence. The suboptimal colouring shown is obtained using the classical greedy heuristic (described in Sec.~\ref{sec:greedy-heuristic}) and the optimal is obtained using the IBM CPLEX Optimiser \cite{cplex2009v12}, resulting in 4 and 2 colour changes respectively. This diagram is based on Figure~1 in Streif et al.~\cite{streif2021beating}.\label{fig:bpsp_concept_example_labels}}
\end{figure*}

The paint shop problem (PSP) consists of a set of optimisation problems in the automotive industry relating to painting instances of cars in sequences various colours satisfying certain constraints (a special case of the minimum bijection problem where the graph is a path~\cite{karpinski2002approximability}). The problem was introduced by Epping et al. \cite{epping2004complexity}, where it was shown to be NP-complete in both the number of colours and the number of car instances in a sequence. There are three main variants of the problem that are typically used. These are, listed in order of increasing granularity, the multi-car multi-colour PSP, the multi-car PSP~\cite{yarkoni2021multi}, and the binary PSP (BPSP). Each problem assumes that the order of cars in the sequence is predetermined by some external factor. This is because the sequence order is more importantly optimised with respect to the factory stages before and after the paint shop, which are typically the body shop stage and the trim/chassis/assembly shop stage respectively~\cite{han2003paint, epping2004complexity}. In the multi-car multi-colour PSP, there can be any number of car instances of the same body type in the sequence, where there are colour count requirements for any number of colours for each body type. For example, there could be five body type A cars and three body type B cars in a sequence, where body type A cars require two to be red, two to be blue and one to be green, while body type B cars require two to be red and one to be blue. This version of the problem is general in nature and can be directly used to model a variety of paint shop problems in modern car factories. The multi-car PSP is more specific with the added constraint that there can only be up to two colours per car body type. This version of the problem is directly relevant to factories for optimising the choice of filler coats. The BPSP, shown in Fig.~\ref{fig:bpsp_concept_example_labels}, adds a further constraint where there can only be up to two cars of each body type in the sequence, and they each are required to be painted different colours (e.g. red and blue). 

The BPSP can be formally defined as the following
\begin{definition}[Binary Paint Shop Problem (BPSP)]
    Let $S = \{s_1, \ldots, s_{n}\}$ be a non-empty sequence of car instances, let $B = \{b_1, \ldots, b_m\}$ be a non-empty set of car body types, and let $\Omega : S \rightarrow B$ be a map that labels the car body type for each car instance in $S$. A tuple $(S, B, \Omega)$ is an input to the BPSP when $S$ contains exactly two car instances of each car body type in $B$ under the map $\Omega$. That is, for all $b_k \in B$, there exists exactly two distinct car instances $s_i, s_j \in S$ such that $\Omega (s_i) = \Omega (s_j) = b_k$. Let $\mathcal{C}$ be a set of colours. A \textit{colouring} is a map $f: S \rightarrow \mathcal{C}$ that assigns a colour from $\mathcal{C}$ to each car instance in $S$. Each colouring has a property called the number of colour changes $\Delta_C$, which is the number of neighbouring car instances $s_i, s_{i+1} \in S$ where $f(s_i) \neq f(s_{i+1})$. For the BPSP with input $(S, B, \Omega)$, a colouring $f_2$ maps to a set of two colours $\mathcal{C}_2 = \{1, 2\}$ and has the constraint that each pair of car instances in $S$ that are assigned the same car body type from $B$ under $\Omega$ are assigned different colours in $\mathcal{C}_2$ under $f_2$. That is, for all distinct pairs $s_i, s_j \in S$ such that $\Omega(s_i) = \Omega(s_j)$, it must follow that $f_2(s_i) \neq f_2(s_j)$. The number of colour changes for $f_2$, can be calculated as $\Delta_C = \sum_{i=1}^{n-1} |f_2(s_i) - f_2(s_{i+1})|$ since $|f_2(s_i) - f_2(s_{i+1})| = 1-[f_2(s_i) = f_2(s_{i+1})]$ when $f_2(\cdot) \in \{1, 2\}$, where $[\alpha]$ is the Iverson bracket of the statement $\alpha$ (1 when true, 0 otherwise). The BPSP can now be stated as the following. Given an input $(S, B, \Omega)$, find a colouring $f_2$ that minimises the number of colour changes $\Delta_C$, that is 
    \begin{align}
         f_2^* &:= \argmin_{f_2}\sum_{i=1}^{n-1} |f_2(s_i) - f_2(s_{i+1})|,\\
         \Delta^*_C &:= \sum_{i=1}^{n-1} |f_2^*(s_i) - f_2^*(s_{i+1})|,
    \end{align}
    where $f_2^*$ and $\Delta^*_C$ denote an optimal colouring and corresponding minimised colour change count respectively.
\end{definition}

Even though the BPSP is simpler than the multi-car variants of the problem, it is still NP-complete and APX-hard \cite{bonsma2006complexity, meunier2009paintshop}, making it an interesting problem for analysis that could lead to new ideas for the more general problems.

\begin{exmp}[BPSP]
Let $(S, B, \Omega)$ be the input for an 8-car BPSP instance, that is, there is a sequence of eight cars $S$ that are each assigned a car body from a set $B$ such that two copies of each body are present in the sequence. The sequence of cars $S$ and set of car bodies $B$ can be written as
\begin{align*}
    S &= \{s_1, s_2, s_3, s_4, s_5, s_6, s_7, s_8\}, \\
    B &= \{{\color{dark_grey}\circlesymbol}, {\color{dark_grey}\trianglesymbol}, {\color{dark_grey}\squaresymbol}, {\color{dark_grey}\pentagonsymbol} \}.
\end{align*}
The car bodies of the cars in the sequence are defined through the relation $\Omega$ as
\begin{align*}
     \Omega(s_2) &= \Omega(s_7) = {\color{dark_grey}\circlesymbol}, \\
     \Omega(s_1) &= \Omega(s_3) = {\color{dark_grey}\trianglesymbol}, \\
     \Omega(s_5) &= \Omega(s_8) = {\color{dark_grey}\squaresymbol}, \\
     \Omega(s_4) &= \Omega(s_6) = {\color{dark_grey}\pentagonsymbol}.
\end{align*}
The sequence of car bodies is used to summarise the problem
\begin{equation*}
    \Omega(S) = \{{\color{dark_grey}\trianglesymbol}, {\color{dark_grey}\circlesymbol}, {\color{dark_grey}\trianglesymbol}, {\color{dark_grey}\pentagonsymbol}, {\color{dark_grey}\squaresymbol}, {\color{dark_grey}\pentagonsymbol}, {\color{dark_grey}\circlesymbol}, {\color{dark_grey}\squaresymbol}\}.
\end{equation*}
The goal is to find a colouring $f_2: S \rightarrow \mathcal{C}_2$, where $\mathcal{C}_2 = \{1, 2\} \equiv \{\text{\color{red}R}, \text{\color{blue}B}\}$, that minimises the number of colour changes~$\Delta_C$ such that the colour of cars that have the same body are different, that is,
\begin{align*}
    & &f_2(s_2) \neq f_2(s_7), & &f_2(s_1) \neq f_2(s_3),& & \\
    & &f_2(s_5) \neq f_2(s_8), & &f_2(s_4) \neq f_2(s_6).& &
\end{align*}
There are many approaches to this problem. An example solution from a classical greedy heuristic (described in Sec.~\ref{sec:greedy-heuristic}) is
\begin{align*}
    & &f^\prime_2(S) = \{\text{\color{red}R}, \text{\color{red}R}, 
    \text{\color{blue}B}, 
    \text{\color{blue}B}, 
    \text{\color{blue}B}, 
    \text{\color{red}R}, 
    \text{\color{blue}B}, 
    \text{\color{red}R}\}, & &\Delta^\prime_C = 4. & &
\end{align*}
The optimal solution can be obtained using an exact solver such as CPLEX~\cite{cplex2009v12}, resulting in 
\begin{align*}
    & &f^*_2(S) = \{\text{\color{red}R}, \text{\color{blue}B}, 
    \text{\color{blue}B}, 
    \text{\color{blue}B}, 
    \text{\color{blue}B}, 
    \text{\color{red}R}, 
    \text{\color{red}R}, 
    \text{\color{red}R}\}, & &\Delta^*_C = 2. & &
\end{align*}
A succinct presentation can be used to display both the problem input and the solution by labelling the car instances by their bodies and drawing them with their solution colours,
\begin{equation}
    \{{\color{red}\trianglesymbol}, {\color{blue}\circlesymbol}, {\color{blue}\trianglesymbol}, {\color{blue}\pentagonsymbol}, {\color{blue}\squaresymbol}, {\color{red}\pentagonsymbol}, {\color{red}\circlesymbol}, {\color{red}\squaresymbol}\}.
\end{equation}
\end{exmp}

\subsection{Mapping the BPSP to QAOA}\label{sec:map-bpsp-to-qaoa}
To solve combinatorial optimisation problems using QAOA, they are first transformed into Ising Hamiltonian energy minimisation problems. These Hamiltonians have already been obtained in previous studies for the BPSP~\cite{streif2021beating} and the multi-car PSP~\cite{yarkoni2021multi}. We describe the transformation for the BPSP here with emphasis on careful justification for the construction of the Ising Hamiltonian. Remarkably, it is possible to map the BPSP to QAOA using a single qubit for each car body type, rather than for each car instance. With such a mapping, the state of each qubit corresponds to the colour assigned to the first occurrence of the two car instances with the same corresponding body type. Couplings between qubits in the Ising Hamiltonian are determined by iterating through the sequence of car instances and observing the relational positions of the car body types. For each adjacent pair of car instances in the sequence, if the pair are both first or both second occurrences of their assigned body type, then they can be influenced to have the same colour by adding a coupling between their corresponding qubits with a $-1$ coupling strength. If one car instance of the pair is the first occurrence of their assigned body type while the other is the second occurrence, then the cars can be influenced to have the same colour by adding a +1 coupling between their corresponding qubits. In the case that both car instances of the adjacent pair have the same body type, then the term becomes a constant since $Z_iZ_i = 1$. Finally, a value of 1/2 is added to each term in the Ising Hamiltonian's sum over couplings so that the energy corresponds to the number of colour changes. After following these steps and refactoring the expression to sum over pairs of adjacent body types, we obtain a BPSP Ising Hamiltonian that directly represents an Ising graph
\begin{equation}
    H_{\text{BPSP}} = \frac{1}{2}\sum_{(i,j)\in E} J_{ij} Z_i Z_j + \frac{C}{2},\label{eq:bpsp_hamiltonian_constant}
\end{equation}
where $J_{ij}$ is the total coupling strength between the pair of qubits corresponding to body types $i$ and $j$, $E$ are the edges in the Ising Hamiltonian's corresponding Ising graph with non-zero coupling, and $C$ specifies an integer constant. A detailed derivation of this expression is included in App.~\ref{sec:bpsp-hamiltonian-derivation}.

The Ising graph approaches 4-regular and $J_{ij} = \pm 1$ couplings with the distribution $P(J_{ij} = -1) = 2/3$ and $P(J_{ij} = 1) = 1/3$ as the problem size increases~\cite{streif2021beating}. Note that the BPSP Ising Hamiltonian is in the same form as the Hamiltonian for integer weighted MAX-CUT
\begin{align}
    H_{\text{MaxCut}} &= -\frac{1}{2} \sum_{(i,j)\in E} J_{ij}(1 - Z_i Z_j) \\
    &= \frac{1}{2}\sum_{(i,j)\in E} J_{ij}Z_i Z_j + \frac{C^\prime}{2},
\end{align}
where $C^\prime$ is an integer constant different to $C$ from Eq.~\eqref{eq:bpsp_hamiltonian_constant}.

\subsection{The QAOA Depth and its Limitations}\label{sec:qaoa-depth-limitations}
When choosing the QAOA depth $p$, there is a trade-off to consider. Higher values of $p$ reduce the error caused by the time discretisation of the Trotterised approximation, however higher values of $p$ also require a higher number of quantum circuit gates to implement, which increases the amount of noise in the computation. There is also an algorithmic performance limitation that occurs when $p$ is too low with respect to the number of qubits $N$. The QAOA Hamiltonian is local, meaning that it is a sum of operators that are bounded in the number of qubits that they act on, which for QAOA is two qubits. Due to this limitation, measured qubits of a QAOA ansatz at QAOA depth $p$ can only be correlated with qubits that are within the $p$-neighbourhood of it, which includes qubits within $p$ distance from the measured qubit in the graph representing the problem Ising Hamiltonian. If the $p$-neighbourhood does not span all of the qubits, then there can be bounds on the algorithmic performance. In Bravyi \textit{et al.}~\cite{bravyi2020obstacles} it was shown that there exist problem graphs $G$ such that when~$p < (\log_2(N)/3-4)/D$, where~$D \geq 3$ is the maximum degree of vertices in~$G$, the QAOA variational energy for~$Z_2$-symmetric problem Hamiltonians (such as for MAX-CUT) on~$G$ is upper bounded by~$(5/6 + \sqrt{D-1}/3D)E_\text{max}$, where~$E_\text{max}$ is the exact largest energy of the problem Hamiltonian. For more typical instances of the max independent set problem, it was shown in~\cite{farhi2020quantum} that when qubits have an average degree of $d$, the~QAOA solution approximation is bounded away from optimal on average by a constant for fixed~$d$ when~$p$ is less than a small degree-dependent multiple of~$\log_2(N)$. In a follow-up paper~\cite{farhi2020quantumWorst}, a similar result was shown for $d$-regular bipartite graphs for the MAX-CUT and max independent set problems. In separate studies, it was shown that the performance of constant depth QAOA for bounded-degree problem Ising graphs cannot achieve the same scaling as the best classical algorithms for increasing problem sizes~\cite{basso2022performance, hastings2019classical}. However, these QAOA depth limitation studies conclude that when the $p$-neighbourhood for each qubit covers the whole problem graph, then there is no indication that the algorithm's power is limited.

\subsection{The Recursive Quantum Approximate Optimisation Algorithm (RQAOA)}
\begin{figure*}
     \centering
     \includegraphics[width=0.80\linewidth]{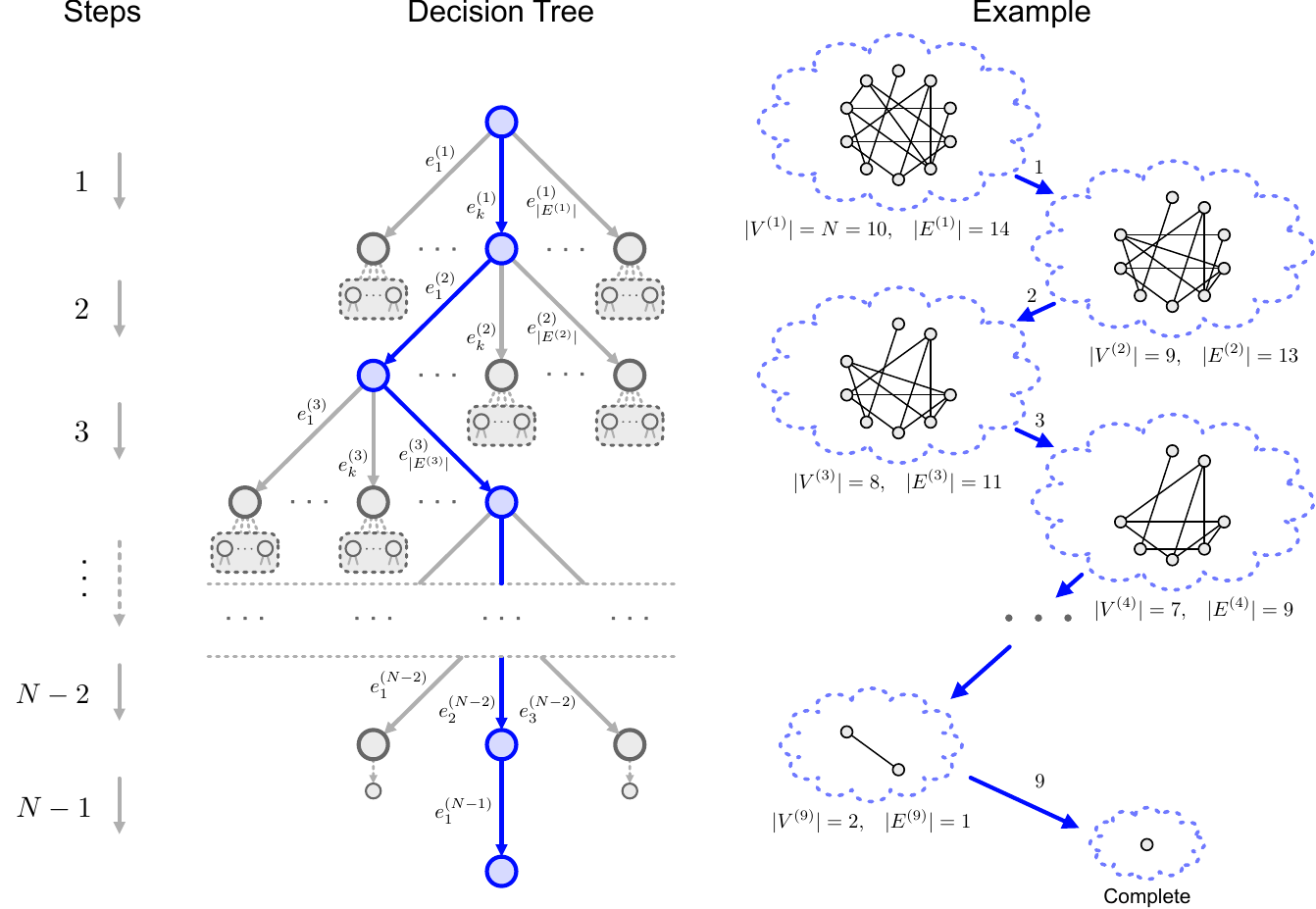}
     \caption[]{A graphic highlighting the RQAOA reduction process. Eliminating variables by choosing edge correlations to round at each RQAOA reduction step can be thought of as conducting a classical greedy depth-first search on a decision tree. The traversed direction at each step is chosen by rounding the $ZZ$ correlation of the pair of qubits with the largest correlation magnitude, a depth-first search heuristic where the $ZZ$ correlations are calculated using a quantum device. By rounding a $ZZ$ correlation to $\pm 1$, a logical relation is established between the qubits, enabling one of them to be removed from the problem Ising Hamiltonian and assigned a state classically once the reduction process completes. The available choice of edges to round,~$e_k^{(i)}$ in the diagram, are the edges in the reduced Ising graph during step $i$, where $k \in \{1, 2, \ldots |E^{(i)}|\}$, and the number of nodes and edges in the Ising graph during step $i$ are denoted by $|V^{(i)}|$ and $|E^{(i)}|$ respectively. There are many paths in the decision tree that lead to optimal solutions. For instance, let $\Gamma$ be a distribution of problem instances in which all states have an equal likelihood of being solution states when selecting a problem instance at random from $\Gamma$. Given a random problem instance from $\Gamma$, during each RQAOA step, if an edge is assigned a $ZZ$ correlation rounding of $\pm 1$ with a uniform probability distribution, then there is a 50\% probability that the choice is correct, leading the path to progress towards a solution state (ignoring solution degeneracy).\label{fig:rqaoa_tree}}
\end{figure*}

The RQAOA was introduced by Bravyi \textit{et al.}~\cite{bravyi2020obstacles} in 2019 to overcome the above limitations relating to the QAOA Hamiltonian being local. It was shown to outperform standard QAOA for frustrated Ising models on random 3-regular graphs. It was later shown analytically for $p=1$ to produce an approximation ratio of 1 for MAX-CUT on $2n$-vertex complete graphs~\cite{bae2024recursive}, while baseline QAOA's is strictly less than $1 - 1/(8n^2)$. An overview diagram of RQAOA is provided in Fig.~\ref{fig:rqaoa_tree}. It works by initially performing QAOA and optimising the ansatz state with respect to the parameters as usual. Then the correlations between each of the adjacent qubits in the problem graph are calculated as $M_{i,j} = \langle \beta, \gamma| Z_i Z_j |\beta, \gamma\rangle$. The largest magnitude correlation is chosen and a constraint is imposed on it to effectively round it to $\pm 1$,
\begin{equation}
    Z_j = \text{sgn}(M_{i,j}) Z_i.
\end{equation}
Substituting this into the problem Hamiltonian allows one of the variables to be eliminated, resulting in a smaller problem graph. The incident edges of the nodes corresponding to the two variables are merged as a result, with a minus sign multiplied to the edge weights depending on the relation. Then QAOA is applied again to this smaller graph to find the largest magnitude correlation and the process is repeated until the problem graph is reduced to be small enough to classically solve exactly. This process typically requires $N-1$ steps to fully reduce the problem graph to a single node, where $N$ is the number of car bodies. In some cases, a single step will decrease the reduced graph node count by more than 1. This is because when merging incident edges of logically related variables that are incident to the same node, the two edges combine into one with a resulting weight that can possibly cancel to zero. This will effectively remove the edge from the graph, which can possibly cause a single qubit to become isolated from the rest of the graph, enabling it to be removed.

The RQAOA can be thought of as a greedy tree search, where directions are chosen based on measured correlations. Using the largest correlation as the direction is only a simple method for making a decision. Significant improvements can be achieved by implementing more advanced methods. This has been demonstrated, by using reinforcement learning~\cite{patel2024reinforcement}, problem-specific classical subroutines~\cite{finvzgar2024quantum}, quantum random access codes~\cite{kondo2025recursive}, and successive interference cancellation combined with majority vote~\cite{gulbahar2025majority}.

\begin{figure*}
     \centering
     \includegraphics[scale=1]{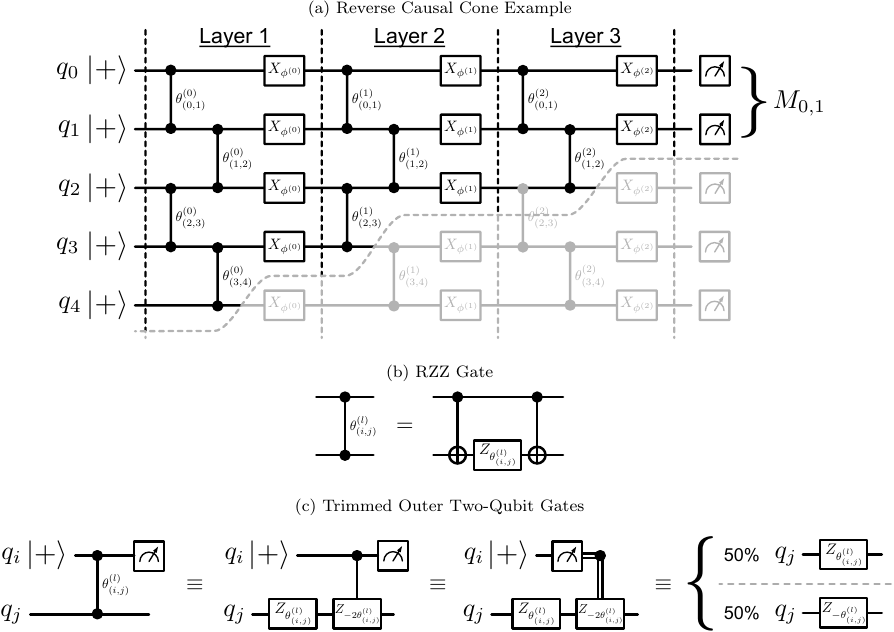}
     \caption[]{\label{fig:RCC}(a) An example of the reverse causal cone (RCC) for a line Ising graph QAOA circuit. The $ZZ$ correlation $M_{0,1}$ only requires qubits 0 and 1 to be measured. Only the gates that contribute to the qubit 0 and 1 measurements are needed. The greyed out gates can be removed since they can be iteratively commuted to the end of the circuit. (b) The shorthand definition used for $R_{ZZ}(\theta_{(i,j)}^{(l)})$ gates applied to qubits $i$ and $j$ within QAOA layer $l$. (c) Circuits showing the equivalence of single-qubit noise gates replacing trimmed two-qubit gates introduced in Section~\ref{sec:trimmed-rcc}. The final noise gate has $R_{Z}(\theta_{(i,j)}^{(l)})$ applied in half of the shots and $R_{Z}(-\theta_{(i,j)}^{(l)})$ applied in the other half.}
\end{figure*}

\subsubsection{Reverse-Causal Cones (RCCs)}
The RQAOA allows us to avoid needing to prepare the full ansatz state. This is because RQAOA only needs to know the largest correlation magnitude between each adjacent pair of qubits to round a correlation to $\pm 1$ when reducing the graph and the parameterised ansatz state can be optimised at each reduction step by calculating the energy at each optimisation iteration, which is just a weighted sum of the correlations. When measuring a correlation, only the operations that connect to the pair of qubits within their $(p-l+1)$-neighbourhood for each QAOA layer $l$ can influence their measured state. An example is shown in Fig.~\ref{fig:RCC}. The set of these operations for a target pair of qubits $(i, j)$ is called the reverse-causal cone (RCC) and is the set of gates not commuting with the operator product $Z_i Z_j$, which can be determined by working backwards through the ansatz state preparation circuit~\cite{streif2020training}. The process for constructing it is the following. In the last QAOA layer $l=p$, include all two-qubit operators and involved qubits that include the target pair of qubits. In the second last layer $l=p-1$, include all two-qubit operators that involve any qubit that has been included so far. Repeat this procedure until all layers have been processed. The number of RCC circuits used to calculate the energy of an Ising graph is equal to its edge count, which for an $N$-node regular graph, scales as~$\mathcal{O}(N)$. The total number of RCC circuits for RQAOA is then 
\begin{equation}
    \text{(RCC circuit count)} = \sum_{i=1}^{N-1} |E^{(i)}|,
\end{equation}
where $|E^{(i)}|$ is the number of edges in the Ising graph during reduction step $i$. The number of edges monotonically decreases with respect to graph reduction, thus the total number of RCC circuits for RQAOA on a regular problem graph scales as $\mathcal{O}(N^2)$.

\subsubsection{Trimming RCC Outer Two-Qubit Gates}\label{sec:trimmed-rcc}
In short-depth QAOA circuits, there are often cases when the RCC's first layer uses extra qubits that are not included in the second layer. These extra qubits and the~$R_{ZZ}(\theta)$ gates that connect them can be trimmed and replaced by equivalent single-qubit noise gates applied to the corresponding connected qubits, as shown in Figure~\ref{fig:RCC}c. The noise gates can be implemented by splitting the circuit shots into two halves which each apply either the $R_Z(\theta)$ or $R_Z(-\theta)$ gate based on the equal probability of measuring a~0 or~1 on the trimmed neighbour qubit before the~$R_{ZZ}(\theta)$ gate is applied. This means that by distributing the shots equally across each possibility of~0 and~1, the two-qubit gates and corresponding qubits can be replaced with their corresponding single-qubit $R_Z(\pm\theta)$ gates. These trimmed RCC circuits are generally easier to implement and have lower levels of gate error, along with a lower quantum resource cost overhead when compiling to hardware. 

RCC circuits with multiple trimmed qubits can be implemented in practice by constructing a circuit for each bitstring that represents the possible measured states of the removed qubits. Each of these trimmed circuits have corresponding $R_Z(\pm\theta)$ gates applied based on their associated removed qubit bit value in the bitstring. Using this approach, the number of unique trimmed RCC circuits, $\Gamma$, used to implement an RCC circuit is upper bounded by $2^k$ where~$k$ is the number of removed qubits. As for the number of removed qubits, given a~$d$-regular problem Ising graph with QAOA depth $p$, the number of qubits in the first layer of an RCC is upper bounded by~$\sum_{l=0}^p 2(d-1)^{l}$, assuming there is the maximum of~$\sum_{l=0}^{p-1} 2(d-1)^{l}$ qubits included in the second layer. The number of removed qubits~$k$ is thus upper bounded as 
\begin{equation}
     k \leq 2(d-1)^{p}.  
\end{equation}
Therefore, the number of unique trimmed RCC circuits~$\Gamma$ is upper bounded as
\begin{equation}
    \Gamma \leq 2^{2(d-1)^{p}}.
\end{equation}
It is also limited when increasing depth~$p$ begins to make the RCC qubits saturate the problem Ising graph, which reduces the possible additional qubits that can be in the first layer and not the second. For an untrimmed~RCC circuit with~$N$ number of shots, we evenly divide the shots between each of the trimmed RCC circuits, resulting in $N/2^k$ shots each. The correlation~$M_{i,j}$ is then calculated as the average over correlations measured from each of the trimmed RCC circuits. An added benefit of dividing the shots in this way is that the variance of the measured correlation is reduced because it no longer depends on the probabilistic nature of measurement outcomes of the outer qubits.

It is tempting to further remove the noise gates entirely, because reducing circuit noise is often the goal in quantum computing. It turns out that the noise-gate choices of $R_Z(\theta)$ or $R_Z(-\theta)$ from a single trimmed qubit have the same effect on the $ZZ$-correlation due to symmetry, and is different to removing the noise gate altogether.  Thus, when there is only a single trimmed qubit, the noise gate can be replaced by only one of its possible choices of gate. For any additional trimmed qubits, the full noise gates are required. This is because different combinations of noise-gate angle choices can induce different correlation effects across qubits that affect the $ZZ$-correlation observable. 

\subsection{Bypassing the QAOA optimisation loop using precomputed parameters}

The QAOA typically involves a classical optimisation loop over the QAOA parameters to search for extrema of the cost function where the cost is evaluated as the QAOA ansatz state energy at each iteration using a quantum device. This optimisation process typically requires many evaluations of the ansatz energy since the volume of parameter space grows exponentially with the QAOA depth $p$~\cite{cook2020quantum} and the energy landscape is prone to vanishing gradients due to barren plateaus, which under certain conditions can exponentially decrease the gradient with respect to qubit count~\cite{holmes2022connecting, mcclean2018barren}. This is particularly a concern in NISQ era since the noisy stochastic nature of quantum devices increases the difficulty in navigating the energy landscape, and barren plateaus can be further induced by noise~\cite{wang2021noise}. 

There are a variety of methods to alleviate the problem of parameter optimisation \cite{alam2020accelerating, wauters2020reinforcement, amosy2024iterative}. It turns out that on families of problem graphs that share certain characteristics, the optimised QAOA parameters concentrate about fixed values~\cite{galda2021transferability, shaydulin2023parameter,lotshaw2021empirical,cook2020quantum, ozaeta2022expectation, wurtz2021fixed, zhou2020quantum}. The mean optimal parameters for random problem instances appear to be independent of problem size and the variance tends to decrease as problem size increases~\cite{streif2020training,zhou2020quantum}. For unweighted MAX-CUT problem graphs, the median QAOA parameters computed over all graphs of the same size, appear to give good approximation ratios for typical instances~\cite{lotshaw2021empirical}. This allows for the possibility of calculating QAOA parameters that can be transferred to unseen instances, enabling a good approximation to be achieved without requiring a classical optimisation loop. The convergence of QAOA parameters and the transferability between different instances can be explained and predicted based on the types of local subgraphs within the problem graph~\cite{galda2021transferability}. By observing the energy landscape of all local subgraphs of certain families of problem graphs (such as all even-regular or odd-regular graphs) separately there are clear patterns that have each subgraph's global extrema overlap in similar regions of the QAOA parameter space. This indicates that certain QAOA parameters will cause the QAOA ansatz energy for all of the local subgraphs to be close to a global minima (or maxima) at the same time, minimising (or maximising) the total QAOA ansatz energy. For MAX-CUT on regular graphs with fixed weights, fixed parameters can be calculated by analysing local subgraphs that are worst-case under a no small cycles conjecture, producing solutions with approximation ratios better than some large worst-case guarantee~\cite{wurtz2021maxcut}. Comparing with the Goemans-Williamson (GW) algorithm on generic graphs~\cite{goemans1995improved}---the classical MAX-CUT polynomial algorithm based on semidefinite programming with the best approximation ratio guarantee $\alpha = 0.87856$---these precomputed QAOA parameters have been shown to give a performance guarantee on 3-regular graphs better than the GW algorithm for $p \geq 11$ with the guarantee approaching unity with a gap proportional to $1/\sqrt{p}$~\cite{wurtz2021fixed}. This performance guarantee also applies to general $d$-regular graphs with approximation ratio proportional to $1/\sqrt{d}$ for fixed $p$. There are also results showing that averaging optimal QAOA parameters over many random 3-regular MAX-CUT instances give parameters that provide good approximations to other larger 3-regular MAX-CUT instances~\cite{brandao2018fixed}. Additionally, a neural network approach has been shown to calculate problem-specific parameters that provide near-optimal approximations for individual MAX-CUT problem graphs~\cite{amosy2024iterative}. Fixed parameters for MAX-CUT on regular graphs can be efficiently calculated in the limit of infinite problem size via a classical variation of QAOA that is based on tensor network techniques called tree-QAOA~\cite{streif2020training}. Furthermore, analytical formulas have been derived to calculate $p=1$ QAOA expectation values for general MAX-CUT graphs~\cite{ozaeta2022expectation, wang2018quantum, hadfield2018quantum} and to further analytically optimise with respect to $p=1$ QAOA parameters for classes of Ising graphs~\cite{ozaeta2022expectation}. The QAOA expected energy for typical instances of the Sherrington-Kirkpatrick (SK) model has also been formulated in the infinite size limit for arbitrary values of $p$ with a time complexity of $\mathcal{O}(16^p)$~\cite{farhi2022quantum}. The parameter transfer methods and results mentioned so far mostly target unweighted MAX-CUT problem graphs. However, it has been shown that median parameters over typical unweighted MAX-CUT instances can be transferred to weighted MAX-CUT problem graphs by rescaling $\boldsymbol{\gamma}$ based on average weight magnitude and average node degree of the target graphs~\cite{shaydulin2023parameter, ozaeta2022expectation}. For a large dataset of random instances with up to 20 nodes and various weight distributions, it was shown that the mean approximation ratio decreased by only 0.02 when using parameter transfer from a single vector (calculated as the median optimised values over all 9-node non-isomorphic unweighted MAX-CUT graphs) of QAOA parameters compared to directly optimising the instances in the dataset. This reduction further improved to only 0.012 when evaluating QAOA using parameter transfer from 10 additional random samples from a pretrained metadistribution of optimised QAOA parameters based on kernel density estimation~\cite{khairy2020learning}. There are also techniques to transfer an arbitrary set of optimal parameters into the adequate domain using symmetries of the problem graph~\cite{lyngfelt2025symmetry}.

There are two main approaches to estimating fixed parameters for the BPSP that we will be using in this work. The first is to use the results for 4-regular graphs with $\pm 1$ couplings obtained in~\cite{streif2021beating} calculated using tensor networks on tree-QAOA~\cite{streif2020training}. The second approach is to estimate the fixed parameters as the median over many smaller random instances using the QAOAKit Python library~\cite{shaydulin2021qaoakit}.

\subsubsection{Approximation Measure}
To compare methods fairly, we aim for a measure of approximation quality that is independent of problem instance and preferably independent of the objective function's properties, such as whether higher values are better or whether the values span positive and negative values. To keep comparisons simple, we construct a measure scaled between 0 and 1 where 1 means that the approximation returns optimal solutions, while 0 means that the approximation is not useful. Using the following definition, also used in \cite{shaydulin2023parameter,villanueva2025hybrid}, the approximation measure for the approximation value is the fraction between the worst case value and the best possible value of the objective function
\begin{equation}
    (\text{approx. measure}) = \frac{(\text{worst}) - (\text{approx. val.})}{(\text{worst}) - (\text{best})}.\label{eq:approx_meas}
\end{equation}
This can also be thought of as one minus the residual energy density~\cite{wauters2020reinforcement}. The ``worst value'' can be specified as the value that is furthest away from the optimal solution, or it can be specified as the average value over inputs that are randomly chosen from a uniform distribution, in which case the approximation measure represents how useful the approximation is compared to random guessing.

\subsection{Classical Approaches}
Here we describe three classical algorithms for solving the BPSP that are used as points of comparison in our analyses: greedy heuristic, recursive greedy heuristic, and exhaustive search. We use the following problem instance as an example to help describe them.
\begin{equation*}
    \Omega(S) = \{{\color{dark_grey}\trianglesymbol}, {\color{dark_grey}\circlesymbol}, {\color{dark_grey}\trianglesymbol}, {\color{dark_grey}\pentagonsymbol}, {\color{dark_grey}\squaresymbol}, {\color{dark_grey}\pentagonsymbol}, {\color{dark_grey}\circlesymbol}, {\color{dark_grey}\squaresymbol}\}.
\end{equation*}

\subsubsection{Greedy Heuristic}\label{sec:greedy-heuristic}
The first car in the sequence is assigned red (which consequently assigns blue to its paired car). Then, iterating through the sequence, each unassigned car is assigned the same colour as the previous car in the sequence. For example,
\begin{align*}
    \Omega(S) &\rightarrow \{{\color{red}\trianglesymbol}, {\color{dark_grey}\circlesymbol}, {\color{blue}\trianglesymbol}, {\color{dark_grey}\pentagonsymbol}, {\color{dark_grey}\squaresymbol}, {\color{dark_grey}\pentagonsymbol}, {\color{dark_grey}\circlesymbol}, {\color{dark_grey}\squaresymbol}\}\\
    &\rightarrow \{{\color{red}\trianglesymbol}, {\color{red}\circlesymbol}, 
    {\color{blue}\trianglesymbol}, {\color{dark_grey}\pentagonsymbol}, {\color{dark_grey}\squaresymbol}, {\color{dark_grey}\pentagonsymbol}, {\color{blue}\circlesymbol}, {\color{dark_grey}\squaresymbol}\}\\
    &\rightarrow \{{\color{red}\trianglesymbol}, {\color{red}\circlesymbol}, 
    {\color{blue}\trianglesymbol}, {\color{blue}\pentagonsymbol}, {\color{dark_grey}\squaresymbol}, {\color{red}\pentagonsymbol}, 
    {\color{blue}\circlesymbol}, {\color{dark_grey}\squaresymbol}\}\\
    &\rightarrow \{{\color{red}\trianglesymbol}, {\color{red}\circlesymbol}, 
    {\color{blue}\trianglesymbol}, {\color{blue}\pentagonsymbol}, 
    {\color{blue}\squaresymbol}, 
    {\color{red}\pentagonsymbol}, 
    {\color{blue}\circlesymbol}, 
    {\color{red}\squaresymbol}\},
\end{align*}
resulting in a colour change count of $\Delta_c=4$.

\subsubsection{Recursive Greedy Heuristic}
Each car body pair is added to the sequence one by one, in the same order they appear in the problem car sequence. Each pair's colours are assigned to minimise the number of colour changes directly after they are added. For example,
\begin{align*}
    \Omega(S) &\rightarrow \{{\color{red}\trianglesymbol}, {\color{blue}\trianglesymbol}\}\\
    &\rightarrow \{{\color{red}\trianglesymbol}, {\color{red}\circlesymbol}, 
    {\color{blue}\trianglesymbol}, {\color{blue}\circlesymbol}\}\\
    &\rightarrow \{{\color{red}\trianglesymbol}, {\color{red}\circlesymbol}, 
    {\color{blue}\trianglesymbol}, {\color{blue}\pentagonsymbol}, {\color{red}\pentagonsymbol}, 
    {\color{blue}\circlesymbol}\}\\
    &\rightarrow \{{\color{red}\trianglesymbol}, {\color{red}\circlesymbol}, 
    {\color{blue}\trianglesymbol}, {\color{blue}\pentagonsymbol}, 
    {\color{red}\squaresymbol}, 
    {\color{red}\pentagonsymbol}, 
    {\color{blue}\circlesymbol}, 
    {\color{blue}\squaresymbol}\},
\end{align*}
resulting in a colour change count of $\Delta_c=3$.

\subsubsection{Exhaustive Search}
This algorithm finds exact solutions to the BPSP by mapping the BPSP instance to an Ising Hamiltonian using the method shown in Sec.~\ref{sec:map-bpsp-to-qaoa} then uses an eigensolver to find the eigenvector with the smallest eigenvalue, which corresponds to the minimum number of colour changes. In this case, the NumPy minimum eigensolver simply sorts the diagonal elements and returns the smallest, essentially performing an exhaustive search. The returned solution is
\begin{align*}
    \Omega(S) &\rightarrow \{{\color{red}\trianglesymbol}, {\color{blue}\circlesymbol}, 
    {\color{blue}\trianglesymbol}, {\color{blue}\pentagonsymbol}, 
    {\color{blue}\squaresymbol}, 
    {\color{red}\pentagonsymbol}, 
    {\color{red}\circlesymbol}, 
    {\color{red}\squaresymbol}\},
\end{align*}
resulting in a colour change count of $\Delta_c=2$. 

\section{Results}

\begin{table*}[t]
    \centering
    \begin{tabular}{c|c|c|c|c|c|c|c|c}
        QAOA Depth & $\beta_1$ & $\gamma_1$ & $\beta_2$ & $\gamma_2$ & $\beta_3$ & $\gamma_3$ & $\beta_4$ & $\gamma_4$ \\ \hline
        1 & $-0.39269$ & $0.52358$ & & & & & & \\
        2 & $-0.53411$ & $0.40784$ & $-0.28296$ & $0.73974$ & & & & \\
        3 & $-0.58794$ & $0.35450$ & $-0.42318$ & $0.65138$ & $-0.22301$ & $0.75426$ & & \\
        4 & $-0.60498$ & $0.31500$ & $-0.47780$ & $0.58754$ & $-0.36127$ & $0.67322$ & $-0.18753$ & $0.77120$
    \end{tabular}
    \caption{The precomputed QAOA parameters for 4-regular graphs with unit couplings, referred to as ``BPSP Precomp'' in Fig.~\ref{fig:algorithm_performance_comparisons}. The data was obtained from Streif et. al.~\cite{streif2021beating}, which uses the approach presented in~\cite{streif2020training}. We include the values here for ease of reference.}
    \label{tab:qaoa_params}
\end{table*}

\begin{figure*}
     \centering
     \includegraphics[width=\linewidth]{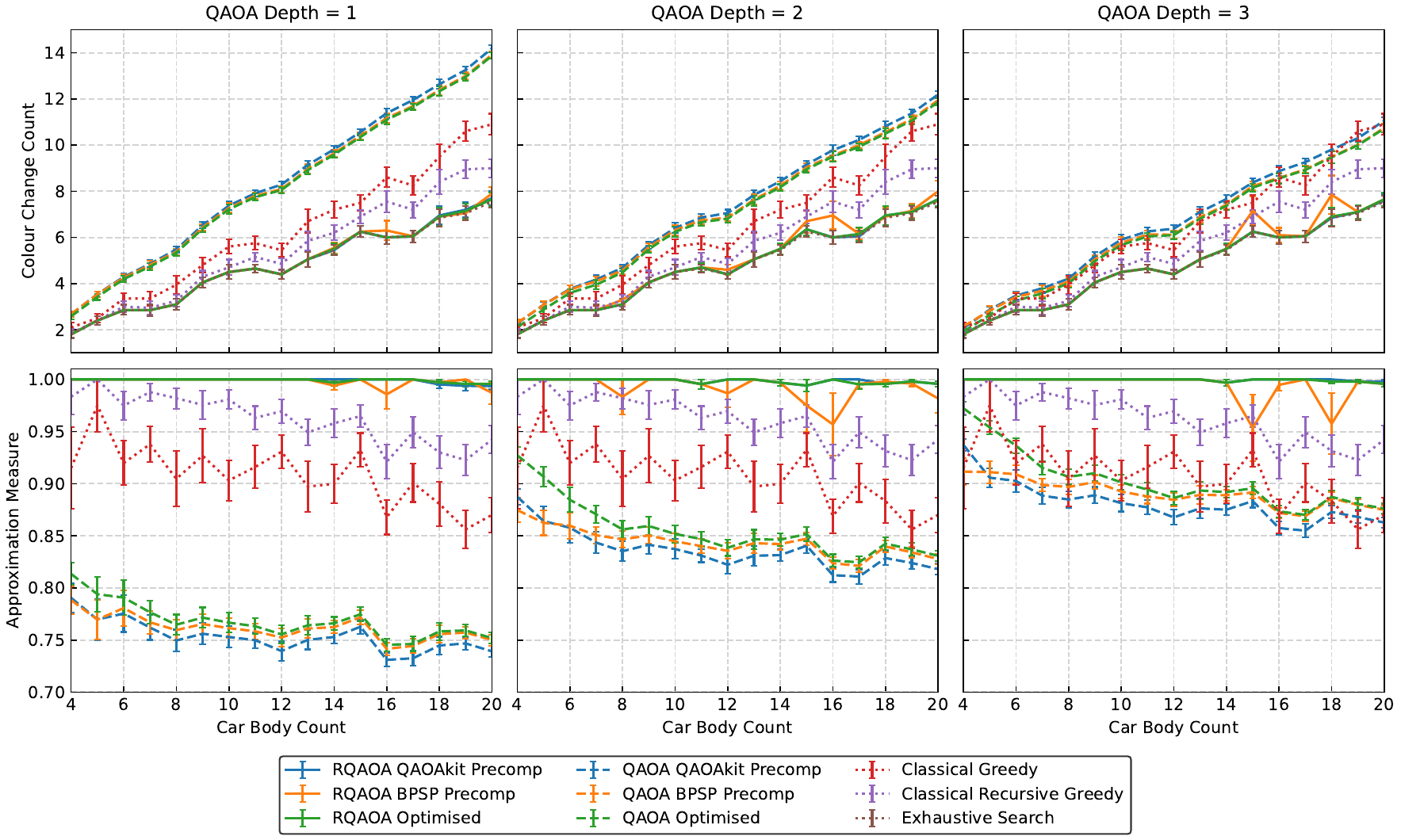}
     \caption[]{The quality of BPSP solutions from a variety of algorithms plotted against the number of car bodies. We compare between~3 approaches to RQAOA,~3 approaches to QAOA and~3 classical algorithms. For RQAOA and QAOA, we show results for optimised parameters as well as parameters obtained using QAOAKit~\cite{shaydulin2021qaoakit}, and parameters obtained by precomputation of 4-regular graphs with unit couplings specifically, from \cite{streif2020training}. The classical algorithms being compared with are the greedy heuristic, recursive greedy heuristic, and the exhaustive search. For each car body count, a set of 20 BPSP instances are chosen at random from a uniform distribution. Each data point is calculated as the mean value obtained from the corresponding set of instances and the error bars are calculated as the standard error. \textbf{(top)} The mean colour change counts for QAOA depths 1, 2, and 3. The RQAOA values shown for the three choices of parameters are almost always equal to the optimal values from the exhaustive search. \textbf{(bottom)} The mean approximation measures for QAOA depths 1, 2, and 3. The approximation measure for a solution of an instance is calculated as the colour change count scaled linearly between the optimal count (corresponding to 1) and the least optimal (corresponding to 0), where the extremas are obtained using exhaustive search.\label{fig:algorithm_performance_comparisons}}
\end{figure*}

\begin{figure*}
     \centering
     \includegraphics[width=0.95\linewidth]{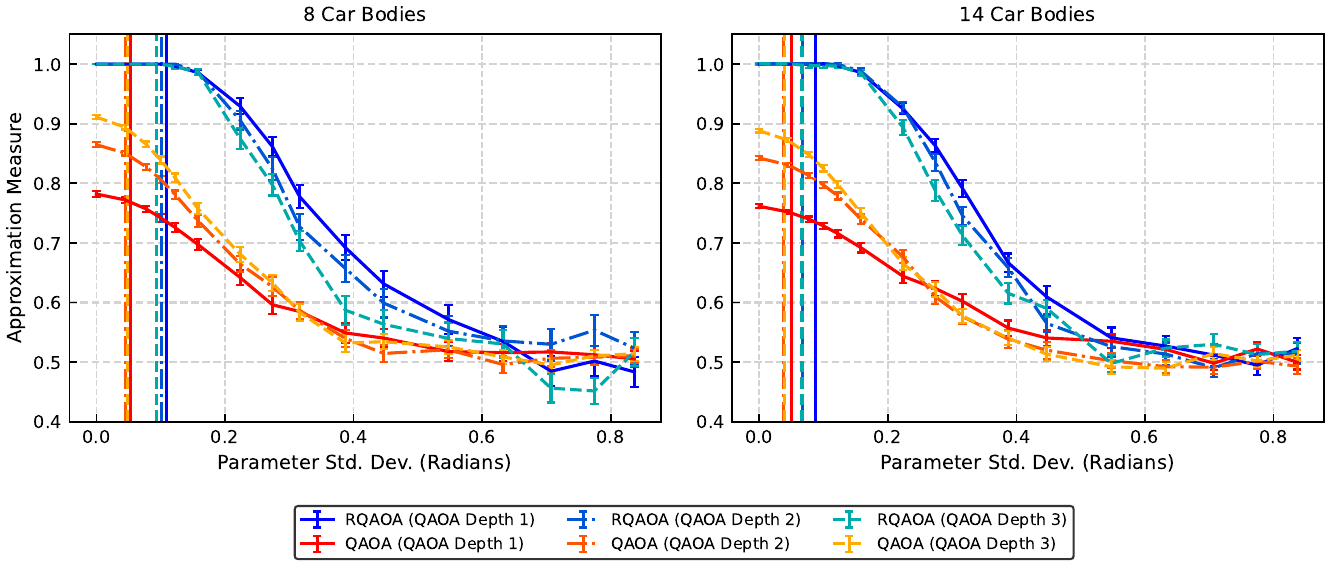}
     \caption[]{Plots showing the effect on performance caused by deviations of QAOA parameters away from optimal. The data are calculated by first obtaining QAOA parameters through optimisation, then adding variation that is sampled from a normal distribution with the specified standard deviation. For RQAOA, parameters are varied at each reduction step by resampling. The vertical lines are the standard deviations of displacements between QAOAKit parameters and optimised, calculated over all instances, and all reduction steps for RQAOA. There are 100 random BPSP instances for each car body count. Each circuit is executed on a statevector simulator with 4096 shots and all optimised data are generated using the Nelder-Mead optimiser with a parameter absolute-error tolerance of~0.0001. All initial parameters for optimisation are obtained using QAOAKit. Error bars are calculated as the standard error.
     \label{fig:param_variance_comparisons}}
\end{figure*}

\subsection{RQAOA Performance with Precomputed QAOA Parameters}
We coded RQAOA in Python and used a local IBM Quantum statevector simulator to obtain results for random BPSP instances, comparing the quality of solutions with QAOA and classical approaches. To maximally observe the effectiveness of the quantum algorithm itself, we do not use a classical solver once the RQAOA is reduced to a certain size (like is suggested in the original paper). Instead, the graph reduction continues until one qubit remains, at which point the solution is obtained through the logical relations recorded at each reduction step. We generate a sample of 20 random BPSP instances for each car body count ranging from 4 to 20. The performance of each approach is measured using two sample averages on the solutions: the colour change count and the approximation measure (described in Eq.~\eqref{eq:approx_meas}). The sample error bars correspond to standard errors. We compare a total of nine approaches to the BPSP, three for RQAOA, three for QAOA and three classical algorithms. For each reduction step of RQAOA and for QAOA, the three methods of choosing the parameters are: optimisation, QAOAKit precomputations, and tailored precomputations for 4-regular unit-coupling graphs (labelled ``BPSP Precomp'' and listed in Table~\ref{tab:qaoa_params}). The three classical algorithms are the greedy heuristic, the recursive greedy heuristic, and the exhaustive search for optimal solutions. Energy evaluations in RQAOA and QAOA are calculated using a statevector simulator with 4096 shots and all optimised data are generated using the Nelder-Mead optimiser with a parameter absolute-error tolerance of 0.0001 and initial QAOA parameters obtained using QAOAKit. 

The performance results are shown in Fig.~\ref{fig:algorithm_performance_comparisons}. We can see that all approaches to choosing parameters for RQAOA perform near-optimally, even for QAOA depth~$p=1$. The RQAOA QAOAKit Precomp performs just as well as QAOAKit Optimised, and occasionally performs marginally better (as shown for 14 car $p=1$ and 17 car $p=2$ approximation measures). There are occasional dips in solution quality when using precomputed BPSP parameters, however overall, it still performs close to optimally. This is surprising, because although the problem graph is initially close to 4-regular with unit-couplings, its structure can drastically change as the graph is reduced during the steps of RQAOA. 

To better understand the effects of unoptimal QAOA parameters, we added variation to optimised parameters to observe the effects on performance, with results shown in Fig.~\ref{fig:param_variance_comparisons}. Using the same simulation settings used for Fig.~\ref{fig:algorithm_performance_comparisons}, The data are calculated for various standard deviations by initially obtaining QAOA parameters through optimisation, then adding a variation that is sampled from a normal distribution with the specified standard deviation. For RQAOA, parameters are varied at each reduction step by resampling. Each data point is averaged over 100 random BPSP instances for each car body count. The vertical lines are the standard deviations of displacements between QAOAKit parameters and optimised, calculated over all instances, and all reduction steps for RQAOA. The data shows that RQAOA is less sensitive to parameter variation than QAOA, indicating a form of noise robustness in the parameters. We suspect that RQAOA also has a level of robustness to quantum device noise as well, however that was not directly tested for in these simulations. The data also shows that the variation caused by QAOAKit is comfortably within the region where RQAOA gives optimal results. As the problem size increases up to 14 car bodies, we observe no notable changes to the performance in relation to the amount of parameter variation. However, the QAOAKit variations for RQAOA, shown by blue vertical lines, appear to decrease for increasing problem sizes. This is likely because the QAOAKit deviation from optimal is largest for small graphs that are highly irregular, while larger BPSP instances have lower proportions of RQAOA reduction steps with Ising graphs that fall in this regime. We suspect that this will eventually begin to increase again for large problem sizes due to the extrapolation precision of QAOAKit being limited by data obtained on relatively small problem graphs. Additionally, for all problem sizes used in these simulations, we observe that RQAOA produces optimal solutions. For large enough problem sizes where RQAOA no longer consistently gives optimal solutions, it is possible that this noise robust region will no longer appear. Although, for such large problems, we believe that the performance of RQAOA with QAOAKit precomputed parameters will remain superior to QAOA, even with optimised parameters.

\begin{figure*}
     \centering
     \includegraphics[width=0.9\linewidth]{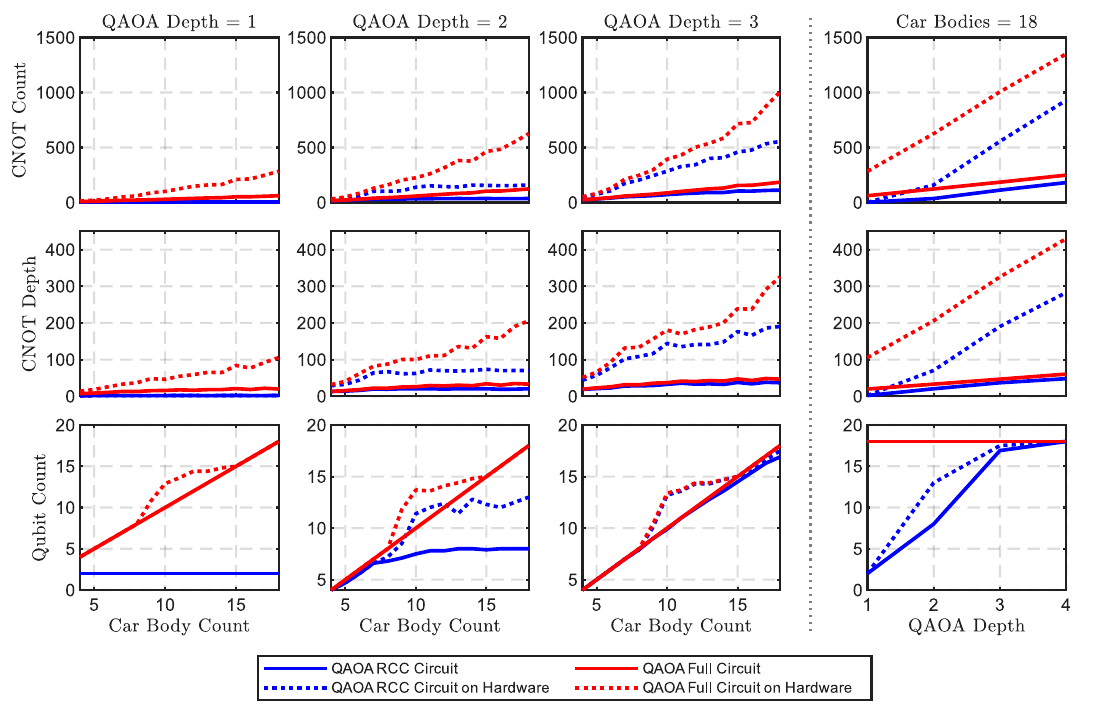}
     \vspace{-4 pt}\caption[]{Quantum resource requirements for QAOA Reverse Causal Cone (RCC) and full circuits. The plots compare the circuit CNOT counts, CNOT depths, and qubit counts for QAOA depths 1 to 4 and car body counts 4 to 18. Each data point is the average over 8 random problem instances. The circuits are mapped to IBM Quantum's heavy-hex hardware using Qiskit's transpiler (v0.26.2) which uses the StochasticSwap mapping algorithm, and any initial SWAP gates are removed.  \label{fig:circuit_resources}}
\end{figure*}

\subsection{Classical and Quantum Resource Requirements}

To better understand how RQAOA scales in terms of quantum resource requirements, we have recorded a variety of resource metrics and compiled the results. 

In Fig.~\ref{fig:circuit_resources}, we compare the CNOT counts, CNOT depths, and qubit counts within QAOA RCC and full circuits, for QAOA depths 1 to 4 and problem sizes ranging from 4 to 18. In particular, we compare the trimmed RCC circuits against the full circuits. Additionally, we compare metrics for circuits before and after compilation to the 27-qubit \textit{ibmq\_montreal} device using Qiskit's transpiler (v0.26.2) with the StochasticSwap hardware mapping algorithm with any initial SWAP gates removed (since qubits can be simply relabelled in code). All data points are calculated as the average over 8 random BPSP instances. The results show that for QAOA depths 1 and 2, the RQAOA RCC circuit resource requirements are substantially lower than the full circuits and are approximately constant with respect to increasing numbers of car bodies. This is likely because the included numbers of qubits in the trimmed RCCs, upper bounded by $\sum_{l=0}^{p-1} 2\times 3^{l}$ (equalling 2 and 8 for $p=1$ and $p=2$ respectively), are much fewer than the problem graph qubit counts. Thus, increasing the size of the problem graph does not have much effect on the span of the RCC. For QAOA depth 3, the number of qubits in the trimmed RCCs is upper bounded by 32 and begins to reliably span the full problem graphs. This leads to the RCC circuit resources growing for increasing car body count throughout our simulations of up to 18 qubits. As the QAOA depth increases, the relative resource differences between RCCs and full circuits appear to diminish, although the reductions in CNOT counts and depths for transpiled circuits remains significant. For higher QAOA depths, we observe that the CNOT count and CNOT depth tends to scale linearly with respect to car body count. Similarly, for 18 car bodies, we notice the CNOT count and CNOT depth scaling linearly with respect to QAOA depths.

\begin{figure*}
     \centering
     \includegraphics[width=0.90\linewidth]{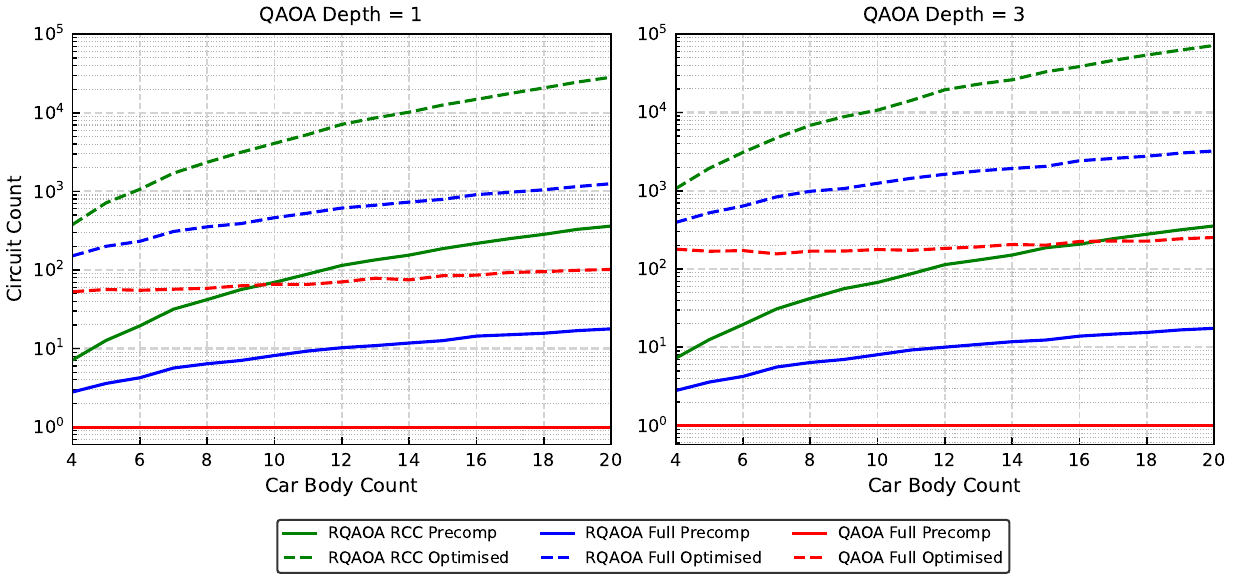}
     \caption[]{The number of circuits used for each quantum algorithm. The algorithms compared are the precomputed-parameter and optimised-parameter implementations of QAOA, RQAOA using Reverse Causal Cones (RCCs), and RQAOA without using RCCs (Full). Each circuit is executed on a statevector simulator with 4096 shots and all optimised data are generated using the Nelder-Mead optimiser with a parameter absolute-error tolerance of~0.0001. All precomputed and initial parameters are obtained using QAOAKit. For each car body count, values are averaged over~20 random BPSP instances with error bars calculated as the standard error. A log-scale y-axis is chosen to visually separate the data, which scales polynomially in the number of car bodies.
     \label{fig:circuit_counts_vs_cars}}
\end{figure*}

\begin{figure*}
     \centering
     \includegraphics[width=0.9\linewidth]{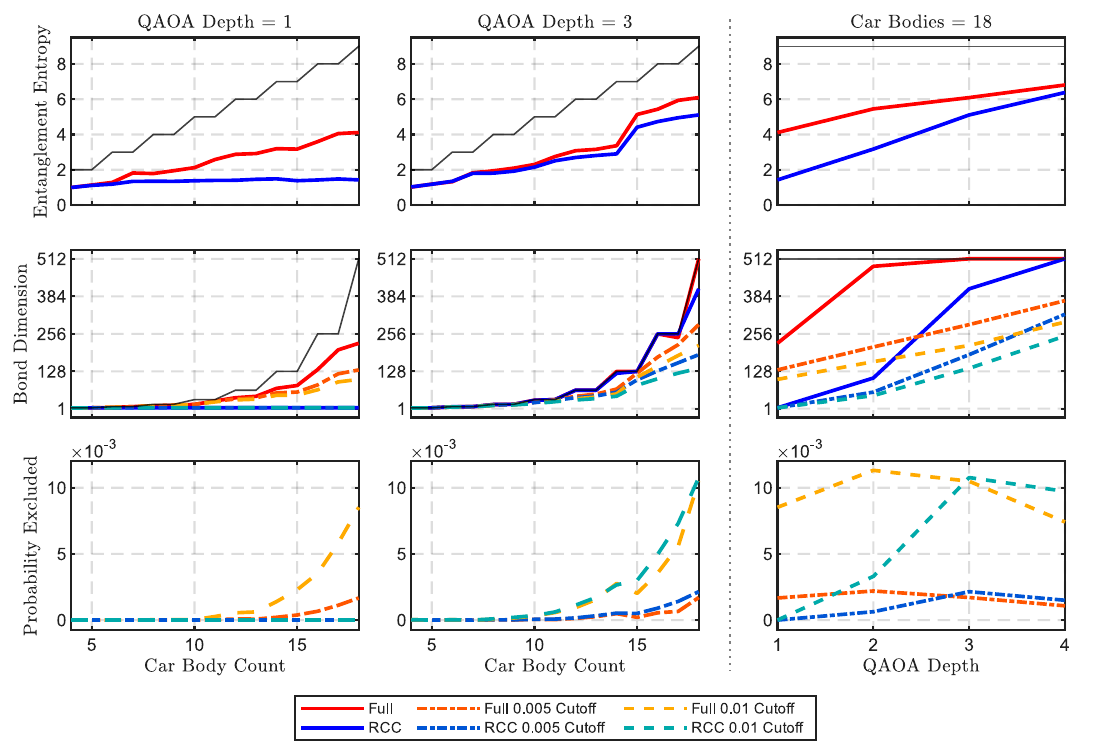}
     \vspace{-4 pt}\caption[]{Resource usages for 1D Matrix Product State (MPS) simulations of QAOA circuits with and without Reverse Causal Cones (RCCs). The names ``Full'' and ``RCC'' in the legend correspond to calculations done on full circuits and circuits split into RCCs respectively. For circuits with RCCs, displayed values are calculated from the RCC with the highest entanglement entropy. The entanglement entropy of the state is recorded as the highest across all contiguous cuts along the MPS line of qubits. The bond dimension is the highest number of non-zero Schmidt coefficients (Schmidt rank) calculated across the cuts. The thin black lines indicate the theoretical maximum values. The cutoff thresholds 0, 0.005, and 0.01 were used to truncate Schmidt coefficients in the Schmidt decomposition of the quantum state. The probability excluded is the total amount of probability removed from the quantum state after truncation. The entanglement entropy variation across cutoff values is considered negligible since the changes peak at approximately 1\%. All values are the average over 8 random BPSP instances.
     \label{fig:circuit_mps_resources}}
\end{figure*}

Further investigating quantum resource requirements, Fig.~\ref{fig:circuit_counts_vs_cars} compares the number of individual circuits required to obtain solutions between the various QAOA-based algorithms. The quantum algorithms presented are the precomputed-parameter and optimised-parameter implementations of QAOA, RQAOA using RCCs, and RQAOA without using RCCs (Full). Choosing to optimise the 2$p$ parameters instead of precompute them introduces a multiplicative factor to the number of circuits that is similar in quantity to the circuit counts shown for QAOA Full Optimised. By assuming precomputed parameters to ignore this overhead, the typical circuit counts for $2N$-car BPSP instances scales as $N-1$ using RQAOA Full, $\mathcal{O}(N^2)$ using RQAOA with RCCs, and is equal to 1 using QAOA. The circuit counts for full RQAOA with precomputed parameters are expected to exceed the circuit counts for fixed 20-qubit optimised QAOA at minimum qubit counts of 100, 200, and 400 for QAOA depths 1, 2, and 3 respectively. Considering that circuit counts for optimised QAOA also appears to be increasing with respect to the qubit count, we expect that circuit counts for full RQAOA with precomputed parameters will always be favourable over optimised QAOA in practice. We find that RQAOA using RCCs and precomputed parameters requires higher numbers of circuits than QAOA with optimisation. This provides a trade-off between the number of quantum circuits and their individual resource requirements. Note that the RCC circuits for this data omit the outer two-qubit gate trimming discussed in Sec.~\ref{sec:trimmed-rcc}.

With the use of RCCs, low-depth RQAOA circuits exceed the performance of full QAOA while being considerably smaller in size. A natural question to ask here is how does the classical simulatability of quantum RCC circuits compare to full QAOA circuits?  We compare the resource usages within classically simulated QAOA circuits in the form of entanglement entropies and bond dimensions of Matrix Product States (MPS) representing quantum states~\cite{dang2019optimising,orus2014practical, schollwock2011density, vidal2003efficient}. By efficiently encoding a state's entanglement structure between qubits, MPS can substantially reduce both storage requirements and computational runtime of quantum algorithm simulations. An MPS expresses a quantum state as a compressed one-dimensional product of tensors, where each tensor corresponds to a single qubit. Neighbouring qubits are connected by an internal bond index representing shared quantum correlations. A pure state can be transformed into MPS form by performing a series of Schmidt decompositions on contiguous bipartitions along the qubit line, each implemented via a singular value decomposition (SVD). Dimensions of the bond indices associated with singular values of zero (or below some cutoff threshold) are removed since they do not significantly contribute to the state. The \textit{bond dimension}, defined as the maximum number of remaining dimensions (Schmidt rank) among all bonds, quantifies the maximal information or entanglement capacity across any cut of the line of qubits and directly determines the computational resources required for MPS simulation. The \textit{entanglement entropy} for a bipartition $A|B$ is given by the von Neumann entropy,
\begin{equation}
    S(A|B) = -\sum_\alpha \lambda^2_\alpha \log(\lambda_\alpha^2),
\end{equation}
where $\lambda_\alpha$ are the non-zero singular values obtained in the~SVD between partitions.

Figure~\ref{fig:circuit_mps_resources} shows these measures computed on full and RCC QAOA circuits of depths 1 to 4 on up to 18 car bodies, averaged over 8 random BPSP instances. The entanglement entropy indicates the level of quantum entanglement required to represent the state, reported as the highest von Neumann entropy across contiguous cuts of the MPS (and across each RCC in RCC circuits). The bond dimension reflects the corresponding classical memory cost and is given by the largest Schmidt rank among the cuts. To investigate the trade-off between MPS precision and classical resource usage, we introduced cutoff thresholds of 0, 0.005, and 0.01 when truncating singular values, and recorded the total probability weight excluded, calculated as the sum of discarded squared singular values summed across all bonds. For reference, the theoretical maximum values of entanglement entropy and bond dimension along a one-dimensional MPS are $\lfloor N/2 \rfloor$ and $2^{\lfloor N/2 \rfloor}$ respectively (9 and 512 for 18 car bodies), shown as thin black lines in the plots. The results show that for QAOA depth 1, while the bond dimension appears to increase exponentially for the Full circuits with respect to car body count, it remains constant when using RCC circuits. The entanglement entropy appears to increase linearly for the Full circuits and is roughly constant when using RCC circuits. As the QAOA depth increases, the curves for Full and RCC circuits appear to converge. This is expected since the number of qubits spanned by RCCs approaches the number of car bodies, $N$, as the depth increases. As the cutoff gets larger, the bond dimensions measured in our data reduce considerably, especially for RCC circuits, while the total probabilities excluded from quantum states remain relatively small. The truncation of singular values induces an approximation error that is upper bounded by two times the total discarded probability $\mathcal{E}$~\cite{schollwock2011density}, implying a pure state fidelity bound of $F\geq 1-\mathcal{E}$.
For these small values of excluded probability, the quantum state maintains a high state fidelity above~$0.99$ for almost all instances, which would have at most marginal effects on the performance of RQAOA. The variation of entanglement entropy across cutoff thresholds is considered to be negligible since the average difference in values peak at approximately 1\%. Therefore, it appears that the combination of RCC circuits and bond dimension cutoffs together help to significantly reduce classical memory costs while maintaining high levels of entanglement entropy and quantum state fidelity. Although the changes in entanglement entropies and state probabilities are small in our data, they will continue to increase as the problem sizes grow. We expect that to maintain small deviations, the cutoff threshold will need to increase to compensate. It would be interesting to see how these results continue to scale with increasing cutoff thresholds and problem sizes. In particular, as the excluded probabilities become larger, how are the RQAOA solution qualities affected? This could help to understand the benefits of combining RCCs and MPS cutoffs for classical simulations of RQAOA and potentially lead to an increased capacity for simulating larger problem sizes.

\section{Conclusion}
We implemented baseline RQAOA in conjunction with precomputed parameters to investigate the impact on performance for BPSP instances compared to using optimised parameters at every RQAOA reduction step. We demonstrated the substantial improvement that RQAOA provides over QAOA, producing near-optimal solutions over all BPSP instances generated up to sequences of 40 cars consisting of 20 car bodies, which are mapped to 20-qubit problem Ising graphs. We found that RQAOA with precomputed parameters from the QAOAKit Python package performed effectively the same whether they were subsequently optimised or not. To further investigate this, we added variations to optimised parameters and recorded their effect on the quality of solutions for BPSPs with up to 14 car bodies. We found that RQAOA exhibits a level of noise robustness in QAOA parameters. We suspect this extends to robustness in the noise of quantum devices as well, however we leave this as an open question for future research. From this data, we confirmed that the standard deviations of displacements between QAOAKit precomputed parameters and optimised parameters are comfortably within the region of tolerance where RQAOA produces its best solutions. 

Additionally, we performed a variety of comparisons between classical and quantum resource requirements, including circuit counts, quantum matrix product state bond dimensions and entanglement entropies, and circuit resources such as qubit counts, CNOT counts, and CNOT depths. In particular, trimmed reverse-causal cone (RCC) circuits, used for measuring $ZZ$-correlations in RQAOA, were compared with full circuits, with consideration for transpilation onto the 27-qubit \textit{ibmq\_montreal} device. We observe that RCCs substantially reduce the circuit resources for QAOA depths 1 and 2 on up to 18 car body BPSP instances, while the benefits begin to diminish as the QAOA depth increases, although the reductions in CNOT counts and depths for transpiled circuits remain significant. For the quantum circuit counts, we find that RQAOA with precomputed parameters typically requires a number of circuits equal to the number of car bodies, which is significantly lower than what QAOA requires to optimise parameters using the Nelder-Mead optimiser. We expect this relation will continue to hold for increasing problem sizes because the circuit counts for QAOA also appears to grow. We find that RQAOA using RCCs and precomputed parameters requires higher numbers of circuits than QAOA with optimisation, hence providing a trade-off between the number of quantum circuits and their individual resource requirements. For bond dimensions and entanglement entropies, we find that with QAOA depth 1, they increase exponentially and linearly respectively for full QAOA circuits, however both remain roughly constant for individual RCCs. As the QAOA depth increases, the curves for full and RCC circuits appear to converge.

Our simulations demonstrate that bypassing the optimisation loop using RQAOA with precomputed parameters is an effective approach to substantially improve solution qualities over optimised QAOA while also significantly reducing circuit counts. Since RQAOA appears to perform well despite its reduction steps significantly modifying the form of the original problem graphs, it is promising that this approach will continue to outperform QAOA on non-BPSP Ising graphs as well.

\medskip
\textbf{Acknowledgements} \par
This work was supported by the University of Melbourne through the establishment of an IBM Quantum Network Hub at the University. CDH was supported by a research grant from the Laby Foundation. The authors gratefully acknowledge the support of the University of Melbourne’s Zero Emission Energy Laboratory (ZEE Lab) and the Victorian Higher Education State Investment Fund (VHESIF).

\medskip
\textbf{Data Availability Statement}
The datasets generated and/or analysed during the current study are available from the corresponding author on reasonable request.

\medskip

\section*{Appendix}
\appendix
\renewcommand\thefigure{\thesection.\arabic{figure}}
\setcounter{figure}{0} 
\renewcommand\thedefinition{\thesection.\arabic{definition}}
\setcounter{definition}{0} 
\renewcommand\thelemma{\thesection.\arabic{lemma}}
\setcounter{lemma}{0}
\renewcommand\thetheorem{\thesection.\arabic{theorem}}
\setcounter{theorem}{0}
\renewcommand\thecorollary{\thesection.\arabic{corollary}}
\setcounter{corollary}{0}

\section{Derivation of QAOA and its Implementation}\label{sec:qaoa-derivation}
Here we give a detailed overview of QAOA and a derivation indicated by the original paper~\cite{farhi2014quantum} to help intuitively understand how it works.

The problem Ising Hamiltonian encoding a QUBO problem can be written as
\begin{equation}
    H_\text{prob} = \sum_{(j,k)} J_{jk} Z_j Z_k + \sum_{j} h_j Z_j,\label{eq:problem-hamiltonian}
\end{equation}
where $Z_j$ is the Pauli-$Z$ operator on qubit~$j$,~$h_j$ is the local field strength on qubit~$j$,~$J_{jk}$ is the coupling strength between qubits~$j$ and~$k$ (where negative values correspond to ferromagnetic couplings), and the first sum is over all pairs of qubits~$(j,k)$. The $\pm 1$ eigenvalues of $Z_j$ operators correspond to the binary values in the QUBO problem instance via $b_j=(1-Z_j)/2$.

The QAOA ansatz state on $N$ qubits is prepared as
\begin{equation}
    |\boldsymbol{\beta}, \boldsymbol{\gamma}\rangle := U(\boldsymbol{\beta}, \boldsymbol{\gamma})|+\rangle^{\otimes N},\label{eq:qaoa-state}
\end{equation}
where $U(\boldsymbol{\beta}, \boldsymbol{\gamma})$ is the QAOA time-evolution operator which is in the form of two alternating unitary operators corresponding to the problem Hamiltonian and a mixer Hamiltonian
\begin{align}
    U(\boldsymbol{\beta}, \boldsymbol{\gamma}) &:= U(\beta_1, \gamma_1, \beta_2, \gamma_2, \ldots \beta_p, \gamma_p) \\
    &= \prod_{l=1}^p U_\text{mix}(\beta_l) U_\text{prob}(\gamma_l),\label{eq:qaoa}
\end{align}
where $p$ is the number of QAOA layers, $U_\text{mix}(\beta)$ is a mixer Hamiltonian time-evolution operator (mixer operator), and $U_\text{prob}(\gamma)$ is the problem Hamiltonian time evolution-operator (phase operator). 

The QAOA ansatz state can be thought of as being constructed from a time discretisation and Trotterised approximation of the time evolution operator for the time-dependent Hamiltonian used in quantum annealing~(QA)~\cite{farhi2000quantum}
\begin{equation}
    H(t) = b(t)H_{\text{mix}} + c(t)H_{\text{prob}},
\end{equation}
where $b,c:[0,T]\rightarrow [0, 1]$ are monotonic functions, where~$b$ is decreasing from $1$ to $0$ and $c$ is increasing from $0$ to~$1$. By the adiabatic theorem of quantum mechanics, if the system evolves slowly enough from an initial ground state of the system (which is $H_{\text{mix}}$ at $t=0$), the system will remain in the ground state~\cite{born1928beweis, kato1950adiabatic}. The mixer Hamiltonian $H_\text{mix}$ is chosen to be 
\begin{equation}
    H_\text{mix} = -\sum_j X_j, 
\end{equation}
such that the initial state $|+\rangle^{\otimes N}$ is its ground state.
The system time-ordered evolution operator can be written as
\begin{align}
    U(t) &= \mathcal{T}\exp\left(-i\int_0^{t} H(t^\prime) dt^\prime\right)\\
    &= \mathcal{T}\exp\left(-i\int_0^{t} \left[b(t^\prime) H_{\text{mix}} + c(t^\prime)H_{\text{prob}} \right] dt^\prime\right),
\end{align}
where $\mathcal{T}$ is the time-ordering operator. The time evolution can be divided into $p+1$ small discrete time steps, $\Delta t := T/(p+1)$, with the assumption that the Hamiltonian is approximately constant during each step. This gives the relation
\begin{equation}
    U(t + \Delta t) \approx e^{-iH(t)\Delta t} U(t),
\end{equation}
which can be iterated backwards through time to give
\begin{align}
    U(T) & \approx e^{-iH(T-\Delta t)\Delta t} e^{-iH(T-2\Delta t)\Delta t} \ldots e^{-iH(0)\Delta t}\\
    &= \prod_{l=1}^{p+1} e^{-i\left[b_l H_{\text{mix}} + c_l H_{\text{prob}}\right]\Delta t}
\end{align}
where $b_l := b\left((p+1-l)\Delta t\right)$ and $c_l := c\left((p+1-l)\Delta t\right)$. Applying the Trotter formula then results in
\begin{align}
    U(T) &\approx \prod_{l=1}^{p+1} \left[e^{-i \Delta t b_l H_{\text{mix}}}e^{-i \Delta t c_lH_{\text{prob}}} \right].
\end{align}
By making the parameter substitutions $\beta_l := -b_l\Delta t$ and $\gamma_l := c_l\Delta t$, and noting that the $l=p+1$ layer can be omitted since the initial state that it acts on is its ground state, we arrive at an expression that matches the form of the QAOA time evolution operator shown in Eq.~\eqref{eq:qaoa}
\begin{align}
    U(\boldsymbol{\beta}, \boldsymbol{\gamma}) &= \prod_{l=1}^p \left[e^{i \beta_l H_{\text{mix}}}e^{-i  \gamma_l H_{\text{prob}}} \right].
\end{align}
Here, the mixer operator is
\begin{align}
    U_\text{mix}(\beta) &= e^{i \beta H_\text{mix}}\\
    &= \prod_{j} e^{-i \beta X_j},
\end{align}
where $\beta \in [0, \pi]$ due to symmetry, and the phase operator is
\begin{align}
    U_\text{prob}(\gamma) &= e^{-i \gamma H_\text{prob}} \\
    &=\left( \prod_{j} e^{-i \gamma h_j Z_j}\right) \left( \prod_{(j,k)} e^{-i \gamma J_{jk} Z_j Z_k}\right),
\end{align}
where $\gamma \in [0, 2\pi]$. 

The QAOA ansatz state $|\boldsymbol{\beta}, \boldsymbol{\gamma}\rangle$ defined in Eq.~\eqref{eq:qaoa-state} can be prepared using an ansatz circuit (or QAOA circuit) by implementing the operators using the following quantum gate representations. Operators in the form of $\exp({-i\alpha X_j/2})$ and $\exp({-i\alpha Z_j/2})$ can be performed using the single-qubit rotation gates $R_x^j(\alpha)$ and $R_z^j(\alpha)$ respectively where~$x$ and~$z$ are the Pauli axes of rotation,~$j$ is the target qubit, and $\alpha$ is the angle of rotation. Operators in the form of $\exp({-i\alpha Z_j Z_k/2})$ can be performed by applying the sequence of gates~$\text{CNOT}^j_k . R_z^k(\alpha) . \text{CNOT}^j_k$ where $\text{CNOT}^j_k$ is a controlled-not gate with control qubit~$j$ and target qubit~$k$.

\section{Refactoring the BPSP Ising Hamiltonian}\label{sec:bpsp-hamiltonian-derivation}
By following the process in Sec.~\ref{sec:map-bpsp-to-qaoa}, we obtain the BPSP Ising Hamiltonian as
\begin{equation}
    H_{\text{BPSP}} = \frac{1}{2}\sum_{i = 1}^{|S| - 1} \left(J^\prime_{s_i,s_{i+1}} Z_{\Omega(s_i)} Z_{\Omega(s_{i+1})} + 1\right),
\end{equation}
where $S$ is the sequence of car instances, $J^\prime_{s_i,s_{i+1}}$ is the coupling strength assigned by the relation between the adjacent cars~$s_i, s_{i+1} \in S$,~$\Omega(s_i)$ is the body type of the car instance~$s_i$, and~$Z_{\Omega(s_i)}$ is the Pauli-$Z$ operator on the qubit that corresponds to body type~$\Omega(s_i)$. In order to obtain the Ising Hamiltonian shown in Eq.~\eqref{eq:bpsp_hamiltonian_constant} that directly represents an Ising graph, the above expression can be simplified by factoring the operator pairs $Z_{\Omega(s_i)} Z_{\Omega(s_{i+1})}$ in the sum such that the sum is over the pairs of adjacent body types $b_i\in B$ in the sequence, instead of the adjacent car instances. The steps for this simplification are as follows, noting that~$[c]$ is the Iverson bracket of the statement $c$ (1 when true, 0 otherwise),
\begin{widetext}
\begin{align}
    H_{\text{BPSP}} &= \frac{1}{2}\sum_{k = 1}^{|B|}\sum_{j = 1}^{k} \left(\sum_{i=1}^{|S|-1} J^\prime_{s_i,s_{i+1}} [(\Omega(s_i), \Omega(s_{i+1})) = (b_j, b_k)] \right) Z_{b_j} Z_{b_k} + \frac{|S| - 1}{2}\label{eq:bpsp_hamiltonian_factor}\\
     &= \frac{1}{2}\left(\sum_{(i,j)\in E} J_{ij} Z_{i} Z_{j} + \sum_{k = 1}^{|B|} [\exists l \in \mathbb{N}, (\Omega(s_l), \Omega(s_{l+1})) = (b_k, b_k)] \right) + \frac{|S| - 1}{2}\label{eq:bpsp_hamiltonian_relabel_indices}\\
     &= \frac{1}{2}\sum_{(i,j)\in E} J_{ij} Z_{i} Z_{j} + \frac{A + |S| - 1}{2}\\
    &= \frac{1}{2}\sum_{(i,j)\in E} J_{ij} Z_i Z_j + \frac{C}{2},
\end{align}
\end{widetext}
where $B$ is the set of body types, $b_i$ is the body type corresponding to qubit $i$, $s_i$ is the $i^\text{th}$ car instance in the sequence~$S$, $J^\prime_{s_i,s_{i+1}}$ is the coupling strength assigned by the relation between the adjacent cars~$s_i, s_{i+1} \in S$, $J_{ij}$ is the total coupling strength between the pair of qubits corresponding to body types $i$ and $j$, $E$ are the edges in the Ising Hamiltonian's corresponding Ising graph with non-zero coupling, $A$ is the number of adjacent pairs of cars in the sequence $S$ with the same body type, $\mathbb{N}$ denotes the set of natural numbers, and $C=A+|S|-1$ specifies an integer constant. Equation~\eqref{eq:bpsp_hamiltonian_relabel_indices} is obtained by summing over the set of edges in the Ising graph while separating the $j=k$ terms (where adjacent cars have the same body type) and relabelling the indices such that~$i$ and~$j$ refer to body types that correspond to nodes in an Ising graph. 

\medskip

\bibliographystyle{unsrt}

\begin{thebibliography}{10}

\bibitem{preskill2018quantum}
John Preskill.
\newblock Quantum computing in the {NISQ} era and beyond.
\newblock {\em Quantum}, 2:79, 2018.

\bibitem{Brandhofer2022}
Sebastian Brandhofer, Daniel Braun, Vanessa Dehn, et~al.
\newblock Benchmarking the performance of portfolio optimization with {QAOA}.
\newblock {\em Quantum Information Processing}, 22(1):25, 2022.

\bibitem{Khumalo2022}
Maxine~T Khumalo, Hazel~A Chieza, Krupa Prag, and Matthew Woolway.
\newblock {An investigation of IBM quantum computing device performance on combinatorial optimisation problems}.
\newblock {\em Neural Computing and Applications}, pages 1--16, 2022.

\bibitem{PhysRevA.104.012403}
Michael Streif, Sheir Yarkoni, Andrea Skolik, et~al.
\newblock Beating classical heuristics for the binary paint shop problem with the quantum approximate optimization algorithm.
\newblock {\em Physical Review A}, 104(1):012403, 2021.

\bibitem{Barkoutsos_2020}
Panagiotis~Kl Barkoutsos, Giacomo Nannicini, Anton Robert, et~al.
\newblock Improving variational quantum optimization using {CVaR}.
\newblock {\em Quantum}, 4:256, 2020.

\bibitem{farhi2014quantum}
Edward Farhi, Jeffrey Goldstone, and Sam Gutmann.
\newblock A quantum approximate optimization algorithm.
\newblock {\em arXiv preprint arXiv:1411.4028}, 2014.

\bibitem{blekos2024review}
Kostas Blekos, Dean Brand, Andrea Ceschini, Chiao-Hui Chou, Rui-Hao Li, Komal Pandya, and Alessandro Summer.
\newblock A review on quantum approximate optimization algorithm and its variants.
\newblock {\em Physics Reports}, 1068:1--66, 2024.

\bibitem{abbas2024challenges}
Amira Abbas, Andris Ambainis, Brandon Augustino, Andreas B{\"a}rtschi, Harry Buhrman, Carleton Coffrin, Giorgio Cortiana, Vedran Dunjko, Daniel~J Egger, Bruce~G Elmegreen, et~al.
\newblock Challenges and opportunities in quantum optimization.
\newblock {\em Nature Reviews Physics}, pages 1--18, 2024.

\bibitem{shor1994algorithms}
Peter~W Shor.
\newblock Algorithms for quantum computation: discrete logarithms and factoring.
\newblock In {\em Proceedings 35th annual symposium on foundations of computer science}, pages 124--134. Ieee, 1994.

\bibitem{grover1996fast}
Lov~K Grover.
\newblock A fast quantum mechanical algorithm for database search.
\newblock In {\em Proceedings of the twenty-eighth annual ACM symposium on Theory of computing}, pages 212--219, 1996.

\bibitem{johnson2007experimental}
David~S Johnson, Gregory Gutin, Lyle~A McGeoch, Anders Yeo, Weixiong Zhang, and Alexei Zverovitch.
\newblock Experimental analysis of heuristics for the {ATSP}.
\newblock {\em The traveling salesman problem and its variations}, pages 445--487, 2007.

\bibitem{Applegate2003ChainedLF}
David Applegate, William Cook, and Andr{\'e} Rohe.
\newblock {Chained Lin-Kernighan for large traveling salesman problems}.
\newblock {\em Informs journal on computing}, 15(1):82--92, 2003.

\bibitem{HELSGAUN2000106}
Keld Helsgaun.
\newblock {An effective implementation of the Lin--Kernighan traveling salesman heuristic}.
\newblock {\em European journal of operational research}, 126(1):106--130, 2000.

\bibitem{10.1145/290179.290180}
Sanjeev Arora.
\newblock {Polynomial time approximation schemes for Euclidean traveling salesman and other geometric problems}.
\newblock {\em Journal of the ACM (JACM)}, 45(5):753--782, 1998.

\bibitem{johnson_mcgeoch_1995}
David~S Johnson and Lyle~A Mcgeoch.
\newblock {\em The traveling salesman problem: a case study in local optimization}.
\newblock John Wiley and Sons, 1995.

\bibitem{10.1007/3-540-52846-6_97}
Michael~L Fredman, David~S Johnson, Lyle~A McGeoch, and Gretchen Ostheimer.
\newblock Data structures for traveling salesmen.
\newblock {\em Journal of Algorithms}, 18(3):432--479, 1995.

\bibitem{bravyi2020obstacles}
Sergey Bravyi, Alexander Kliesch, Robert Koenig, and Eugene Tang.
\newblock Obstacles to variational quantum optimization from symmetry protection.
\newblock {\em Physical Review Letters}, 125(26):260505, 2020.

\bibitem{holmes2022connecting}
Zo{\"e} Holmes, Kunal Sharma, Marco Cerezo, and Patrick~J Coles.
\newblock Connecting ansatz expressibility to gradient magnitudes and barren plateaus.
\newblock {\em PRX Quantum}, 3(1):010313, 2022.

\bibitem{wang2021noise}
Samson Wang, Enrico Fontana, Marco Cerezo, Kunal Sharma, Akira Sone, Lukasz Cincio, and Patrick~J Coles.
\newblock Noise-induced barren plateaus in variational quantum algorithms.
\newblock {\em Nature communications}, 12(1):1--11, 2021.

\bibitem{mcclean2018barren}
Jarrod~R McClean, Sergio Boixo, Vadim~N Smelyanskiy, Ryan Babbush, and Hartmut Neven.
\newblock Barren plateaus in quantum neural network training landscapes.
\newblock {\em Nature communications}, 9(1):1--6, 2018.

\bibitem{zhang2018environment}
Rui Zhang.
\newblock Environment-aware production scheduling for paint shops in automobile manufacturing: A multi-objective optimization approach.
\newblock {\em International Journal of Environmental Research and Public Health}, 15(1):32, 2018.

\bibitem{bonsma2006complexity}
P~Bonsma, Th~Epping, and Winfried Hochst{\"a}ttler.
\newblock Complexity results on restricted instances of a paint shop problem for words.
\newblock {\em Discrete Applied Mathematics}, 154(9):1335--1343, 2006.

\bibitem{meunier2009paintshop}
Fr{\'e}d{\'e}ric Meunier and Andr{\'a}s Seb{\H{o}}.
\newblock Paintshop, odd cycles and necklace splitting.
\newblock {\em Discrete Applied Mathematics}, 157(4):780--793, 2009.

\bibitem{garey1974some}
Michael~R Garey, David~S Johnson, and Larry Stockmeyer.
\newblock Some simplified {NP}-complete problems.
\newblock In {\em Proceedings of the sixth annual ACM symposium on Theory of computing}, pages 47--63, 1974.

\bibitem{wurtz2021fixed}
Jonathan Wurtz and Danylo Lykov.
\newblock The fixed angle conjecture for {QAOA} on regular {MaxCut} graphs.
\newblock {\em arXiv preprint arXiv:2107.00677}, 2021.

\bibitem{ozaeta2022expectation}
Asier Ozaeta, Wim van Dam, and Peter~L McMahon.
\newblock Expectation values from the single-layer quantum approximate optimization algorithm on {Ising} problems.
\newblock {\em Quantum Science and Technology}, 7(4):045036, 2022.

\bibitem{streif2021beating}
Michael Streif, Sheir Yarkoni, Andrea Skolik, Florian Neukart, and Martin Leib.
\newblock Beating classical heuristics for the binary paint shop problem with the quantum approximate optimization algorithm.
\newblock {\em Physical Review A}, 104(1):012403, 2021.

\bibitem{streif2020training}
Michael Streif and Martin Leib.
\newblock Training the quantum approximate optimization algorithm without access to a quantum processing unit.
\newblock {\em Quantum Science and Technology}, 5(3):034008, 2020.

\bibitem{shaydulin2021qaoakit}
Ruslan Shaydulin, Kunal Marwaha, Jonathan Wurtz, and Phillip~C Lotshaw.
\newblock {QAOAKit}: A toolkit for reproducible study, application, and verification of the {QAOA}.
\newblock In {\em 2021 IEEE/ACM Second International Workshop on Quantum Computing Software (QCS)}, pages 64--71. IEEE, 2021.

\bibitem{shaydulin2023parameter}
Ruslan Shaydulin, Phillip~C Lotshaw, Jeffrey Larson, James Ostrowski, and Travis~S Humble.
\newblock Parameter transfer for quantum approximate optimization of weighted maxcut.
\newblock {\em ACM Transactions on Quantum Computing}, 4(3):1--15, 2023.

\bibitem{vijendran2025classical}
V~Vijendran, Dax~Enshan Koh, Ping~Koy Lam, and Syed~M Assad.
\newblock Classical and quantum heuristics for the binary paint shop problem.
\newblock {\em arXiv preprint arXiv:2509.15294}, 2025.

\bibitem{peruzzo2014variational}
Alberto Peruzzo, Jarrod McClean, Peter Shadbolt, Man-Hong Yung, Xiao-Qi Zhou, Peter~J Love, Al{\'a}n Aspuru-Guzik, and Jeremy~L O’brien.
\newblock A variational eigenvalue solver on a photonic quantum processor.
\newblock {\em Nature Communications}, 5(1):1--7, 2014.

\bibitem{cplex2009v12}
IBM~ILOG Cplex.
\newblock V12. 1: User’s manual for {CPLEX}.
\newblock {\em International Business Machines Corporation}, 46(53):157, 2009.

\bibitem{karpinski2002approximability}
Marek Karpinski.
\newblock Approximability of the minimum bisection problem: An algorithmic challenge.
\newblock In {\em International Symposium on Mathematical Foundations of Computer Science}, pages 59--67. Springer, 2002.

\bibitem{epping2004complexity}
Th~Epping, Winfried Hochst{\"a}ttler, and Peter Oertel.
\newblock Complexity results on a paint shop problem.
\newblock {\em Discrete Applied Mathematics}, 136(2-3):217--226, 2004.

\bibitem{yarkoni2021multi}
Sheir Yarkoni, Alex Alekseyenko, Michael Streif, David Von~Dollen, Florian Neukart, and Thomas B{\"a}ck.
\newblock Multi-car paint shop optimization with quantum annealing.
\newblock In {\em 2021 IEEE International Conference on Quantum Computing and Engineering (QCE)}, pages 35--41. IEEE, 2021.

\bibitem{han2003paint}
Yong-Hee Han, Chen Zhou, Bert Bras, Leon McGinnis, Carol Carmichael, and PJ~Newcomb.
\newblock Paint line color change reduction in automobile assembly through simulation.
\newblock In {\em Winter Simulation Conference}, volume~2, pages 1204--1209, 2003.

\bibitem{farhi2020quantum}
Edward Farhi, David Gamarnik, and Sam Gutmann.
\newblock The quantum approximate optimization algorithm needs to see the whole graph: A typical case.
\newblock {\em arXiv preprint arXiv:2004.09002}, 2020.

\bibitem{farhi2020quantumWorst}
Edward Farhi, David Gamarnik, and Sam Gutmann.
\newblock The quantum approximate optimization algorithm needs to see the whole graph: Worst case examples.
\newblock {\em arXiv preprint arXiv:2005.08747}, 2020.

\bibitem{basso2022performance}
Joao Basso, David Gamarnik, Song Mei, and Leo Zhou.
\newblock Performance and limitations of the {QAOA} at constant levels on large sparse hypergraphs and spin glass models.
\newblock In {\em 2022 IEEE 63rd Annual Symposium on Foundations of Computer Science (FOCS)}, pages 335--343. IEEE, 2022.

\bibitem{hastings2019classical}
Matthew~B Hastings.
\newblock Classical and quantum bounded depth approximation algorithms.
\newblock {\em arXiv preprint arXiv:1905.07047}, 2019.

\bibitem{bae2024recursive}
Eunok Bae and Soojoon Lee.
\newblock Recursive {QAOA} outperforms the original {QAOA} for the {MAX-CUT} problem on complete graphs.
\newblock {\em Quantum Information Processing}, 23(3):78, 2024.

\bibitem{patel2024reinforcement}
Yash~J Patel, Sofiene Jerbi, Thomas B{\"a}ck, and Vedran Dunjko.
\newblock Reinforcement learning assisted recursive {QAOA}.
\newblock {\em EPJ Quantum Technology}, 11(1):6, 2024.

\bibitem{finvzgar2024quantum}
Jernej~Rudi Fin{\v{z}}gar, Aron Kerschbaumer, Martin~JA Schuetz, Christian~B Mendl, and Helmut~G Katzgraber.
\newblock Quantum-informed recursive optimization algorithms.
\newblock {\em PRX Quantum}, 5(2):020327, 2024.

\bibitem{kondo2025recursive}
Ruho Kondo, Yuki Sato, Rudy Raymond, and Naoki Yamamoto.
\newblock Recursive quantum relaxation for combinatorial optimization problems.
\newblock {\em Quantum}, 9:1594, 2025.

\bibitem{gulbahar2025majority}
Burhan Gulbahar.
\newblock Majority voting with recursive qaoa and cost-restricted uniform sampling for maximum-likelihood detection in massive mimo.
\newblock {\em IEEE Transactions on Wireless Communications}, 2025.

\bibitem{cook2020quantum}
Jeremy Cook, Stephan Eidenbenz, and Andreas B{\"a}rtschi.
\newblock The quantum alternating operator ansatz on maximum k-vertex cover.
\newblock In {\em 2020 IEEE International Conference on Quantum Computing and Engineering (QCE)}, pages 83--92. IEEE, 2020.

\bibitem{alam2020accelerating}
Mahabubul Alam, Abdullah Ash-Saki, and Swaroop Ghosh.
\newblock Accelerating quantum approximate optimization algorithm using machine learning.
\newblock In {\em 2020 Design, Automation \& Test in Europe Conference \& Exhibition (DATE)}, pages 686--689. IEEE, 2020.

\bibitem{wauters2020reinforcement}
Matteo~M Wauters, Emanuele Panizon, Glen~B Mbeng, and Giuseppe~E Santoro.
\newblock Reinforcement-learning-assisted quantum optimization.
\newblock {\em Physical Review Research}, 2(3):033446, 2020.

\bibitem{amosy2024iterative}
Ohad Amosy, Tamuz Danzig, Ohad Lev, Ely Porat, Gal Chechik, and Adi Makmal.
\newblock Iteration-free quantum approximate optimization algorithm using neural networks.
\newblock {\em Quantum Machine Intelligence}, 6(2):38, 2024.

\bibitem{galda2021transferability}
Alexey Galda, Xiaoyuan Liu, Danylo Lykov, Yuri Alexeev, and Ilya Safro.
\newblock Transferability of optimal {QAOA} parameters between random graphs.
\newblock In {\em 2021 IEEE International Conference on Quantum Computing and Engineering (QCE)}, pages 171--180. IEEE, 2021.

\bibitem{lotshaw2021empirical}
Phillip~C Lotshaw, Travis~S Humble, Rebekah Herrman, James Ostrowski, and George Siopsis.
\newblock Empirical performance bounds for quantum approximate optimization.
\newblock {\em Quantum Information Processing}, 20(12):1--32, 2021.

\bibitem{zhou2020quantum}
Leo Zhou, Sheng-Tao Wang, Soonwon Choi, Hannes Pichler, and Mikhail~D Lukin.
\newblock Quantum approximate optimization algorithm: Performance, mechanism, and implementation on near-term devices.
\newblock {\em Physical Review X}, 10(2):021067, 2020.

\bibitem{wurtz2021maxcut}
Jonathan Wurtz and Peter Love.
\newblock {MaxCut} quantum approximate optimization algorithm performance guarantees for p> 1.
\newblock {\em Physical Review A}, 103(4):042612, 2021.

\bibitem{goemans1995improved}
Michel~X Goemans and David~P Williamson.
\newblock Improved approximation algorithms for maximum cut and satisfiability problems using semidefinite programming.
\newblock {\em Journal of the ACM (JACM)}, 42(6):1115--1145, 1995.

\bibitem{brandao2018fixed}
Fernando~GSL Brandao, Michael Broughton, Edward Farhi, Sam Gutmann, and Hartmut Neven.
\newblock For fixed control parameters the quantum approximate optimization algorithm's objective function value concentrates for typical instances.
\newblock {\em arXiv preprint arXiv:1812.04170}, 2018.

\bibitem{wang2018quantum}
Zhihui Wang, Stuart Hadfield, Zhang Jiang, and Eleanor~G Rieffel.
\newblock Quantum approximate optimization algorithm for {MaxCut}: A fermionic view.
\newblock {\em Physical Review A}, 97(2):022304, 2018.

\bibitem{hadfield2018quantum}
Stuart~Andrew Hadfield.
\newblock {\em Quantum algorithms for scientific computing and approximate optimization}.
\newblock Columbia University, 2018.

\bibitem{farhi2022quantum}
Edward Farhi, Jeffrey Goldstone, Sam Gutmann, and Leo Zhou.
\newblock The quantum approximate optimization algorithm and the {Sherrington-Kirkpatrick} model at infinite size.
\newblock {\em Quantum}, 6:759, 2022.

\bibitem{khairy2020learning}
Sami Khairy, Ruslan Shaydulin, Lukasz Cincio, Yuri Alexeev, and Prasanna Balaprakash.
\newblock Learning to optimize variational quantum circuits to solve combinatorial problems.
\newblock In {\em Proceedings of the AAAI conference on artificial intelligence}, volume~34, pages 2367--2375, 2020.

\bibitem{lyngfelt2025symmetry}
Isak Lyngfelt and Laura Garc{\'\i}a-{\'A}lvarez.
\newblock Symmetry-informed transferability of optimal parameters in the quantum approximate optimization algorithm.
\newblock {\em Physical Review A}, 111(2):022418, 2025.

\bibitem{villanueva2025hybrid}
Jedwin Villanueva, Gary~J Mooney, Bhaskar~Roy Bardhan, Joydip Ghosh, Charles~D Hill, and Lloyd~CL Hollenberg.
\newblock Hybrid quantum optimization in the context of minimizing traffic congestion.
\newblock {\em arXiv preprint arXiv:2504.08275}, 2025.

\bibitem{dang2019optimising}
Aidan Dang, Charles~D Hill, and Lloyd~CL Hollenberg.
\newblock Optimising matrix product state simulations of shor's algorithm.
\newblock {\em Quantum}, 3:116, 2019.

\bibitem{orus2014practical}
Rom{\'a}n Or{\'u}s.
\newblock A practical introduction to tensor networks: Matrix product states and projected entangled pair states.
\newblock {\em Annals of physics}, 349:117--158, 2014.

\bibitem{schollwock2011density}
Ulrich Schollw{\"o}ck.
\newblock The density-matrix renormalization group in the age of matrix product states.
\newblock {\em Annals of physics}, 326(1):96--192, 2011.

\bibitem{vidal2003efficient}
Guifr{\'e} Vidal.
\newblock Efficient classical simulation of slightly entangled quantum computations.
\newblock {\em Physical review letters}, 91(14):147902, 2003.

\bibitem{farhi2000quantum}
Edward Farhi, Jeffrey Goldstone, Sam Gutmann, and Michael Sipser.
\newblock Quantum computation by adiabatic evolution.
\newblock {\em arXiv preprint quant-ph/0001106}, 2000.

\bibitem{born1928beweis}
Max Born and Vladimir Fock.
\newblock Beweis des adiabatensatzes.
\newblock {\em Zeitschrift f{\"u}r Physik}, 51(3):165--180, 1928.

\bibitem{kato1950adiabatic}
Tosio Kato.
\newblock On the adiabatic theorem of quantum mechanics.
\newblock {\em Journal of the Physical Society of Japan}, 5(6):435--439, 1950.

\end{thebibliography}

\end{document}